\pdfoutput=1
\documentclass[12pt]{iopart}
\usepackage[dcucite]{harvard}
\usepackage{algorithm}
\usepackage{algpseudocode}
\usepackage{caption}
\usepackage[labelfont=bf]{caption}
\usepackage{graphicx}
\usepackage{float}
\usepackage{iopams}
\expandafter\let\csname equation*\endcsname\relax
\expandafter\let\csname endequation*\endcsname\relax
\usepackage{amsmath}
\usepackage{algorithm}
\usepackage{algpseudocode}

\usepackage{subfigure}
\usepackage{lipsum}

\usepackage[normalem]{ulem}
\usepackage[usenames,dvipsnames]{color}
 
\newcommand{\rd}{\textcolor{Red}}
\newcommand{\bl}{\textcolor{Blue}}

\newcommand*\diff{\mathop{}\!\mathrm{d}}
\DeclareMathOperator{\sinc}{sinc}
\DeclareMathOperator*{\argmin}{arg\,min}

\begin{document}
{\bf \large Topical Review}



\title[Acoustic inverse problem in PACT]{A survey of computational frameworks for solving the acoustic inverse problem in three-dimensional photoacoustic computed tomography}

\author{[Joemini Poudel$^1$, Yang Lou$^1$] and Mark A Anastasio$^2$}
\address{$^1$Department of Biomedical Engineering, Washington University in St.\ Louis, St. Louis, MO 63130}
\address{$^2$Department of Bioengineering, University of Illinois at Urbana-Champaign, 1406 W Green St., Urbana, IL 61801}

\ead{\mailto{maa@illinois.edu}}

\begin{abstract}
Photoacoustic computed tomography (PACT), also known as optoacoustic tomography,
 is an emerging imaging technique that holds great promise for biomedical imaging.
PACT is a hybrid imaging method that can exploit the strong endogenous contrast of optical methods
along with the high spatial resolution of ultrasound methods.
 In its canonical form that is addressed in this article, PACT seeks to estimate the photoacoustically-induced
 initial pressure distribution within the object.
Image reconstruction methods are employed to solve the acoustic inverse problem associated
with the image formation process.  When an idealized imaging scenario is considered,
analytic solutions to the PACT inverse problem are available; however, in practice,
numerous challenges exist that are more readily addressed within an optimization-based, or iterative,
image reconstruction framework.  In this article, the PACT
image reconstruction problem is reviewed within the context of modern optimization-based  image reconstruction
methodologies.
 Imaging models that relate the measured photoacoustic
 wavefields to the sought-after object function are described in their
 continuous and discrete forms. The basic principles
 of optimization-based image reconstruction from discrete PACT measurement data are presented,
 which includes a review of methods for modeling the PACT measurement system response and
other important physical factors.  Non-conventional formulations of the PACT image 
reconstruction problem, in which acoustic parameters of the medium are concurrently
estimated along with the PACT image, are also introduced and reviewed. 

\end{abstract}
\pacs{87.57.nf, 42.30 wb} 
\submitto {Physics in Medicine and Biology}
\maketitle


\section{Introduction to PACT}
\label{sec:section1}

Photoacoustic computed tomography (PACT), also known as  optoacoustic tomography,
is a rapidly emerging imaging technique that holds great promise for biomedical
imaging  \cite{Kruger:95,Kruger:99,Xu:2002,Xu:2003,OraBook}.
PACT is a hybrid technique that exploits the photoacoustic effect for
signal generation.
When a sufficiently short optical pulse 
is employed to irradiate an object such as biological tissue, the
photoacoustic  effect results in the generation
of acoustic signals within the object.  After propagating out of the 
object, these signals can be measured by use of wide-band ultrasonic transducers.
  The goal of PACT in its canonical formulation is
 to reconstruct an image that represents a map of the initial pressure distribution within the object
from knowledge of the measured photoacoustically-induced acoustic signals.
The initial pressure distribution is proportional to the absorbed optical energy distribution
within the object, which can reveal diagnostically useful information based
on endogenous hemoglobin contrast or exogenous contrast if molecular probes are utilized.
As such, PACT can be viewed either as an ultrasound mediated optical
modality or an ultrasound modality that exploits optical-enhanced image contrast \cite{XuReview}.

Over the past few decades, there have been numerous fundamental studies of photoacoustic imaging of
biological tissue \cite{Oraev:97,Wang:99,Kruger:99,paltauf:1526,Oraevsky:228,maslov:024006,Esen},
 and the development of PACT continues to progress at a tremendous rate
 \cite{anastasio2017special}.
Biomedical applications of PACT include small animal imaging \cite{xia2014small,ma2009multispectral,brecht2009whole} and human breast imaging \cite{andreev2000optoacoustic,manohar2005twente,lin2018clinical,becker2018multispectral,lou2017system},
to name only a few.  For additional information regarding applications of PACT, the reader is referred
to the many review articles that have been published on this topic \cite{Li-Review,beard2011biomedical,WangReview,wang2012photoacoustic,ntziachristos2010molecular}.


PACT is a computed imaging modality
that utilizes an image reconstruction algorithm for image formation.
From a physical perspective, the image reconstruction problem
in PAT corresponds to an acoustic inverse source problem \cite{Anastasio:IP}. When an idealized imaging scenario is considered, a variety of analytic solutions to the
 PACT inverse problem are available \cite{Li-Review,rosenthal2013acoustic,kuchment2011mathematics,agranovsky2007reconstruction}; however, in practice, numerous challenges exist that can limit their applicability.
Alternatively,  optimization-based, or iterative,
image reconstruction methods for PACT provide the opportunity to enhance image quality
by compensating for physical factors, noise, and data-incompleteness.
While such approaches are routinely employed in the broader image reconstruction community,
relatively few research groups have explored such modern reconstruction methods for PACT. Although they can be computationally demanding, the advent and use of modern parallel computing technologies~\cite{wang2013accelerating} have rendered these reconstruction algorithms feasible for many PACT applications.

 In this article, the PACT image reconstruction problem is reviewed
 within the context of modern optimization-based image reconstruction methodologies.
This review is restricted to the  problem of estimating
the initial pressure distribution and does not address
the more complicated problem of recovering the optical
 properties of an object \cite{saratoon2013gradient,yuan2006quantitative}.
Imaging models that relate the measured photoacoustic wavefields to the sought-after object function
 are described in their continuous and discrete forms.
These models will describe 
physical non-idealities in the data such
as those introduced by acoustic inhomogeneity, attenuation, and  the
response of the imaging system.
 The basic principles of optimization-based PACT image
 reconstruction from discrete measurement data are presented, which includes
descriptions of forward operators that accurately describe the physics of image acquisition. Furthermore, the derivation of the adjoints of the corresponding forward operators are also reviewed. The adjoint operators facilitate the application of gradient-based approaches in solving the optimization-based PACT image reconstruction problem.
 Non-conventional formulations of the PACT image reconstruction problem, in which acoustic parameters of the medium are concurrently estimated along with the PACT image, are also introduced and reviewed.

The article is organized as follows. In Section~\ref{sec:section2}, two canonical forward models employed for PACT are reviewed that are based on the acoustic wave equation and integral geometry formulations. The explicit formulation of discrete imaging models are described in Section~\ref{sec:section3}. Optimization-based image reconstruction methods that are based on the discrete imaging models are presented in Section~\ref{sec:section4}. Furthermore, joint reconstruction approaches to PACT image reconstruction whereby acoustic parameters of the medium are concurrently estimated along with the initial pressure distribution are discussed in Section~\ref{sec:section5}. Section~\ref{sec:section6} concludes the article by providing a brief overview of the challenges and opportunities for research related to image reconstruction for practical applications of PACT. 

\begin{figure}[h]
  \centering
      {
        \resizebox{3.5in}{!}{
          \includegraphics{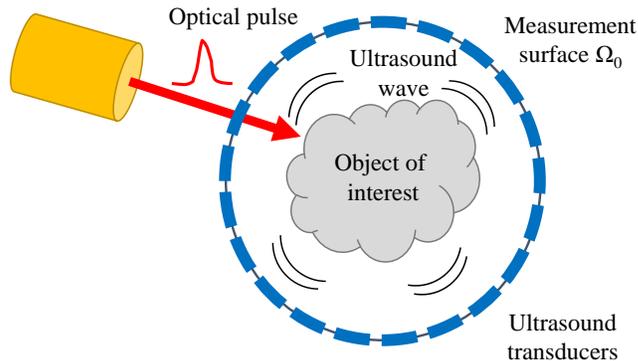}
      }}
      \caption{\label{fig:geom}
A schematic of a PACT imaging experiment. }
\end{figure}

\section{Canonical forward models in their continuous forms}
\label{sec:section2}

A schematic of a general PACT imaging experiment is shown in Fig.~\ref{fig:geom}.
 A sufficiently short~\cite{DieboldBook} laser pulse is employed to irradiate an object at time $t=0$
and the photoacoustic effect results in the generation of
internal pressure distribution inside of the object, which is denoted as $p_0(\mathbf{r})$,
$\mathbf{r} \in \mathbb{R}^3$.
This pressure distribution subsequently propagates outwards in the form of acoustic waves and is detected by the wide-band point-like ultrasonic transducers located on a measurement aperture $\Omega_0 \subset \mathbb{R}^3$. The use of alternative measurement technologies, such as integrating
line or area detectors \cite{burgholzer2005thermoacoustic,paltauf2007photoacoustic}, have also been widely explored but will not be reviewed here.
 The measurement aperture $\Omega_0$ is a two-dimensional (2D) surface
that partially or completely
surrounds the object. The propagated pressure wavefield at time $t > 0$ will be denoted
 as $p(\mathbf{r},t)$.

Below, two types of 3D canonical forward models that relate the measured propagated pressure wavefield to the sought after
object function $p_0(\mathbf{r})$ are described in their continuous forms.  These models,
which form the basis for mathematical studies of the inverse problem in PACT, do not model transducer characteristics.  The effects of finite sampling and other physical factors associated with transducer characteristics
are addressed in the discrete versions of these models presented in Section~\ref{sec:section3}.

\subsection{Wave-equation based PACT forward model in its continuous form}
\label{subsec:2a}

Consider an idealized PACT experiment in which an object with known heterogeneous acoustic properties is irradiated with laser pulse to photoacoustically induce an initial pressure distribution $p_0(\mathbf{r})$. The initial pressure distribution subsequently generates outwardly propagating broadband acoustic waves in the surrounding medium. The acoustic properties of the surrounding medium modulate the behavior of the propagating acoustic waves based on the acoustic wave equation. Let $c_0(\mathbf{r})$ denote the medium's ambient speed of sound (SOS) distribution, $\rho(\mathbf{r},t)$ and $\rho_0(\mathbf{r})$ denote the distributions of the medium's acoustic and ambient densities, respectively. Let $\dot{\mathbf{s}}(\mathbf{r},t)$ denote the particle velocity in the medium. The initial pressure distribution $p_0(\mathbf{r})$ and all quantities that describe the properties of the medium are assumed to be represented by bounded functions that have compact supports.


For a wide range of PACT applications, acoustic absorption is not negligible~\cite{treeby2010photoacoustic,la2006image,burgholzer2007compensation,modgil2012photoacoustic,dean2011effects}. Hence, the strong dependence of the amplitude, spectrum and shape of the propagating broadband acoustic pressure signals on the absorption characteristics of the medium needs to be modeled. It is well known that over diagnostic ultrasound frequency ranges, the acoustic absorption in biological tissue can be described by a frequency power law of the form~\cite{szabo1994time,szabo2004diagnostic} 
\begin{equation}\label{eq:Powerlaw}
\alpha(\mathbf{r},f) = \alpha_0(\mathbf{r}) f^y,
\end{equation}
where $f$ is the temporal frequency in MHz, $\alpha_0(\mathbf{r})$ is the spatially varying frequency-independent absorption coefficient in dB MHz$^{-y}$cm$^{-1}$, and $y$ is the power law exponent that is typically in the range of 0.9-2.0~\cite{szabo2004diagnostic}. Note that classical lossy wave equations predict an absorption that is either frequency independent or proportional to frequency squared~\cite{markham1951absorption}.

To describe the effects of power law absorption and dispersion, formulations of the wave equation that model time-domain fractional derivative operators as convolutions have been proposed~\cite{szabo1994time,szabo2004diagnostic,podlubny1998fractional,moshrefi1998physical}. Numerical implementation of time-domain fractional derivative operators for accurately modeling 3D acoustic wave propagation poses a significant memory burden~\cite{yuan1999simulation}. To overcome this memory burden, some work has focused on devising a recursive algorithm to compute the time-domain fractional derivative~\cite{liebler2004full}. However, this approach is heuristic and requires \textit{a priori} optimization for each value of $y$. 
To circumvent this, a lossy  wave equation modeling power law absorption using fractional Laplacian operators has been proposed~\cite{chen2004fractional}. The fractional Laplacian operators can be easily implemented numerically to effectively model the power law absorption in 3D lossy media. Although, the proposed fractional Laplacian derivative operator-based lossy wave equation can accurately model power law absorption, it does not exhibit the correct dispersive sound speed relation~\cite{sushilov2004frequency}. In order to account for the dispersion inconsistencies, a lossy wave equation that utilizes two  fractional Laplacian derivative operators was proposed~\cite{treeby2010modeling,treeby2010photoacoustic}. These two fractional Laplacian derivative operators account for the required power law absorption and dispersion terms, separately. 

Given the frequency-dependent power law absorption model, one can formulate the PACT forward model in a lossy, acoustically heterogeneous fluid media as~\cite{treeby2010modeling,morse1968theoretical,morse1961linear}:  
\begin{subequations}\label{eq:2a_loss}
\begin{align}
\frac{\partial}{\partial t} \dot{\mathbf{s}}(\mathbf{r},t) &= - \frac{1}{\rho_0(\mathbf{r})} \nabla p(\mathbf{r},t)\\
\frac{\partial}{\partial t} \rho(\mathbf{r},t) &=  - \rho_0(\mathbf{r}) \nabla \cdot \dot{\mathbf{s}}(\mathbf{r},t)\\
p(\mathbf{r},t) 
= c_0(\mathbf{r})^2 &\Big\{ 1 - \mu(\mathbf{r}) \frac{\partial}{\partial t} (-\nabla^2)^{\frac{y}{2} - 1} - \eta(\mathbf{r}) (-\nabla^2)^{\frac{y-1}{2}}\Big\} \rho(\mathbf{r},t),\label{eq:2a_lossiii} \\
\textnormal{subject to the initial conditions}& \nonumber \\
p(\mathbf{r},t)\Big|_{t = 0} &= p_0(\mathbf{r}),\
\ \ \  \dot{\mathbf{s}}(\mathbf{r},t)\Big|_{t= 0}= 0.
\end{align} 
\end{subequations}
Here, the spatially varying quantities $\mu(\mathbf{r})$ and $\eta(\mathbf{r})$ describe the acoustic absorption and dispersion proportionality coefficients that are defined as
\begin{equation}
\mu(\mathbf{r}) = - 2\alpha_0(\mathbf{r}) c_0(\mathbf{r})^{y-1}, \ \ \ \eta(\mathbf{r}) = 2 \alpha_0(\mathbf{r}) c_0(\mathbf{r})^y \tan\Big(\frac{\pi y}{2}\Big).
\end{equation}
The second and third terms on the right hand side of Eqn.~\eref{eq:2a_lossiii} account for the required power law absorption and dispersion terms separately through two fractional Laplacian derivative operators. The initial value problem defined in Eqn.~\eref{eq:2a_loss} describes the propagation of photoacoustically generated pressure data $p(\mathbf{r},t)$ given the spatially varying sound speed $c_0(\mathbf{r})$, ambient density $\rho_0(\mathbf{r})$, and the acoustic absorption coefficient $\alpha_0(\mathbf{r})$.

Consider that the measured pressure data $ p(\mathbf{r},t)|_{\mathbf{r} = \mathbf{r}^{\prime}}$ are recorded outside the support of the object for $\mathbf{r}^{\prime} \in $ $\Omega_0$ and $t \in [0,T]$. The continuous PACT forward model consists of the composition of the partial differential equation (PDE) described in Eqn.~\eref{eq:2a_loss} and an observation operator that restricts the pressure recorded to the measurement surface $\Omega_0$. Hence, the mapping from the initial pressure distribution $p_0(\mathbf{r})$ to the pressure recorded on the continuous measurement aperture $\Omega_0$ in a acoustically heterogeneous fluid media can be expressed as:
\begin{align}\label{eq:2a_HCC}
p(\mathbf{r}^{\prime},t) = \mathcal{H}_{wave} p_0(\mathbf{r}), 
\end{align}
where the wave equation-based forward operator $\mathcal{H}_{wave}:\mathbb{L}_2(\mathbb{R}^3) \mapsto \mathbb{L}_2(\Omega_0) \times [0,T] $ describes a continuous-to-continuous (C-C) mapping
between two function spaces. The reason for the use of such terminology is that the wave equation-based forward operator maps a function of continuous variable (as opposed to a discrete variable) to another function of a continuous variable. There is no implication that either the function itself is continous or even that the mapping is continuous.

\subsection{Integral geometry-based forward model}
\label{subsec:2b}

In the special case of a lossless and acoustically homogeneous infinite medium,
  the solution to the wave equation in Eqn.~\eref{eq:2a_loss} can be expressed as \cite{xu2006photoacoustic}
\begin{equation}
	p(\mathbf{r}', t) = \frac{1}{4 \pi c_0^2} \int_V \diff ^3 \mathbf{r}\; p_0(\mathbf{r}) \frac{d}{dt} \frac{\delta \left( t - \frac{|\mathbf{r}' - \mathbf{r}|}{c_0} \right)}{|\mathbf{r}' - \mathbf{r}|}
	\label{eq:2b_homo_CC} \equiv \mathcal{H}_{int} p_0(\mathbf{r}),
\end{equation}
where $c_0$ denotes the constant sound speed value,
 $\delta(t)$ denotes the one-dimensional (1D) Dirac delta function,
and the integral geometry-based C-C forward operator is denoted as
 $\mathcal{H}_{int}:\mathbb{L}_2(\mathbb{R}^3) \mapsto \mathbb{L}_2(\Omega_0) \times [0,T] $.
 

Equation (\ref{eq:2b_homo_CC}) can be  conveniently expressed in terms of the spherical Radon transform (SRT) \cite{kuchment2011mathematics} as
\begin{equation}
	g(\mathbf{r}', t) = \int_V \diff ^3 \mathbf{r}\; p_0(\mathbf{r}) \delta(c_0 t - |\mathbf{r}' - \mathbf{r}|),
	\label{eq:2b_SRT}
\end{equation}
where the data function $g(\mathbf{r}', t)$ is related to the measured pressure data $p(\mathbf{r}', t)$ by
\begin{equation}
	p(\mathbf{r}', t) = \frac{1}{4 \pi c_0^2} \frac{\partial}{\partial t} \left( \frac{g(\mathbf{r}', t)}{t} \right).
	\label{eq:2b_g_def}
\end{equation}
In the special case of a 2D problem,  the spherical Radon transform model reduces to a circular Radon transform
\cite{haltmeier2007thermoacoustic,wang2008elucidation}.

For the case of a known weakly heterogeneous acoustic media,
 Eqn.\ (\ref{eq:2b_SRT}) can be replaced by  a
 generalized Radon transform model that is derived from the wave equation
by use of a  geometrical acoustics approximation. In that case,
 the measured pressure signals are related to $p_0(\mathbf r)$ through
 integration over nonspherical isochronous surfaces \cite{miller1987new,modgil2010image,jose2012speed}.

 Heuristic strategies have also been proposed  to mitigate image artifacts
that result from employing the idealized forward model in Eqn.\ (\ref{eq:2b_SRT}) in the presence of unknown acoustic heterogeneity.
For example, 
 half-time and partial-time image reconstruction methods have
been proposed for PACT image reconstruction from temporally truncated
measurements that exclude
 components of the measured data that are strongly aberrated \cite{poudel2017mitigation,anastasio2005feasibility,anastasio2005half}.

\section{Discrete forward models}
\label{sec:section3}

\subsection{Review of semidiscrete and discrete forward modeling}
\label{subsec:3a}

As with any digital imaging system, the data acquired in a PACT experiment represent
a finite collection of numbers that form a vector. 
  As such, the PACT forward operator is fundamentally a continuous-to-discrete (C-D)
mapping \cite{barrett2013foundations} that relates $p_0(\mathbf r)$ to the measurement vector.
The C-D operator for PACT can be expressed as
\begin{align}
\mathcal{H}_{CD} \equiv \mathcal{D}\mathcal{H}_{CC},
\end{align}
where $\mathcal{D}$ is a discretization operator that spatially and temporally samples the pressure wavefield $p(\mathbf r^\prime, t)$ and $\mathcal{H}_{CC}$ represents a C-C PACT forward operator such
as $\mathcal{H}_{int}$ or $\mathcal{H}_{wave}$.


Let  ${\mathbf{u}} \in \mathbf{R}^{QK \times 1}$ denote a lexicographically ordered representation of the sampled pressure data and let $[{\mathbf{u}}]_m$ denote its $m$-$th$ component.
When ideal point-like ultrasound transducers are assumed,
the measured samples of the pressure wavefield are given by
\begin{align}
[{\mathbf{u}}]_{qK + k}  = p(\mathbf{r}^{\prime},t)|_{\mathbf{r}^{\prime} = \mathbf{r}_{q}^{\prime}, t = k \Delta_t},
\end{align}
for $k = 0,1,...,K-1$.
 Here, $k$ and $K$ denote the temporal sample index and total number of temporal samples, respectively, and the vectors $\mathbf{r}_{q}^{\prime} \in \mathbb{R}^3$, $q = 0,1,2,...,Q-1$, describe
the locations  of the $Q$ transducers on the aperture $\Omega_0$.

More generally, as discussed in Section~\ref{sec:3d},  the measured pressure data can be described as~\cite{wang2015photoacoustic}
\begin{equation}
\label{eq:CDaperture}
[{\mathbf{u}}]_{qK + k}=[\mathcal{H}_{CD}\,p_0(\mathbf r)]_{qK + k}=
 \int_{-\infty}^\infty\!\! dt\; \tau_{k}(t)
   \int_{\Omega_0}\!\! d \Omega_0 \; p(\mathbf{r}_0,t)
                               \sigma_{q}(\mathbf{r}_0),
\end{equation}
where $\sigma_{q}(\mathbf{r}_0)$ and $\tau_{k}(t)$
are functions that describe the spatial and temporal  sampling apertures of the transducers, respectively.

In order to employ the algebraic or optimization-based image reconstruction methods described later,
the C-D forward operator must be approximated by a discrete-to-discrete (D-D) one.
To accomplish this, a finite dimensional representation of $p_0(\mathbf r)$ can be employed.
An $N$-dimensional representation of $p_0(\mathbf r)$ can
be described as
\begin{equation}\label{D_D:finite_dimensional_representation}
  p_0^a(\mathbf{r})=\sum_{n=1}^{N} [\boldsymbol \theta]_{n}\phi_{n}({\mathbf{r}}),
\end{equation}
where $N>0$ and the superscript $a$ indicates that $p_0^a(\mathbf r)$ is
an approximation of  $p_0(\mathbf r)$.
The functions $\phi_n(\mathbf r)$ are called expansion functions
and the expansion coefficients $[\boldsymbol \theta]_{n}$
are elements of the $N$-dimensional vector $\boldsymbol \theta$.
The goal of image reconstruction is to estimate
$\boldsymbol \theta$ for a fixed choice of the expansion functions $\phi_n(\mathbf r)$.
As described below, different choices for the $\phi_n(\mathbf r)$ and rules for
determining $\boldsymbol \theta$ will result in different D-D forward models.
The specific choice of expansion functions may be motivated by various theoretical and practical reasons including a desire to minimize representation error, incorporation of \emph{a priori} information regarding the object, and efficient computation.
Popular choices of expansion functions in PACT include cubic or radially symmetric expansion functions known as Kaiser-Bessel (KB) window functions \cite{ephrat2008three,paltauf2002iterative,wang2011imaging,wang2011imaging,wang2013accelerating,wang2014discrete}, and linear interpolation functions \cite{zhang2009effects,wang2013accelerating,dean2012accurate,ding2017efficient}.
In general, these D-D imaging models will have distinct numerical properties that will affect the performance of iterative reconstruction
 algorithms \cite{wang2011photoacoustic}.

D-D forward models can be established by substitution of a finite-dimensional object
representation into the C-D imaging model:
\begin{equation}
\label{D_D:dis_to_dis_img_mod}
  {\mathbf{u}} = \mathcal{H}_{CD}\mathbf{p}_0^a(\mathbf r)
   =\sum_{n=1}^N  [\theta]_{n} \mathcal{H}_{CD} \{\phi_{n}(\mathbf{r})\}
   \equiv \mathbf{H}\boldsymbol{\theta},
\end{equation}
where  the D-D operator $\mathbf{H}$ is
commonly referred to as the {\it system matrix}.
An element in the $n$-th row and $m$-th column of $\mathbf{H}$ will be denoted by  $[\mathbf H]_{n,m}$.

Next, examples of D-D PACT imaging models for use with homogeneous and heterogeneous acoustic
media are reviewed.

\subsection{D-D forward models for use with homogeneous media}
\label{subsec:3b}

In this subsection, two popular D-D PACT imaging models for homogeneous acoustic media are reviewed that employ different choices for the expansion functions.
\subsubsection{Interpolation-based D-D PACT model}
\label{subsec:3b_1}
In the interpolation-based D-D PACT imaging model, the the associated expansion functions can be expressed as \cite{kak2001principles}
\begin{equation}
\phi^{\rm int}_n(\mathbf r)\equiv \left\{\begin{array}{ll}
(1-\frac{|x-x_n|}{\Delta_s})
(1-\frac{|y-y_n|}{\Delta_s})
(1-\frac{|z-z_n|}{\Delta_s}), & \text{if}\,
|x-x_n|, |y-y_n|, |z-z_n| \leq  \Delta_s\\
0, & \text{otherwise}
\end{array}\right.,
\label{eq:3b_interpolation_expansion}
\end{equation}
where $\Delta_s$ is the distance between neighboring points on an uniform and isotropic Cartesian grid.  
 The coefficient vector $\boldsymbol{\theta}$ can be defined as \cite{wang2011photoacoustic}: 
\begin{equation}
	[\boldsymbol{\theta}]_n = \int_V \diff ^3 \mathbf{r}\; \delta(\mathbf{r} - \mathbf{r}_n) p_0(\mathbf{r}), \quad n = 0,1,\dots,N-1, 
	\label{eq:3b_int_coef}
\end{equation}
where $\mathbf{r}_n = (x_n, y_n, z_n)^T$ denotes the location of the $n$-th Cartesian grid node.  
The corresponding D-D PACT imaging model based on the interpolation expansion functions can be expressed as
\begin{equation}
	\mathbf{u} = \mathbf{H}_{int} \boldsymbol{\theta}_{int},
	\label{eq:3b_int_DD}
\end{equation}
where
\begin{equation}\label{eq:3b_Hint}
	\mathbf{H}_{int} \equiv \mathbf{D} \mathbf{G}. 
\end{equation}
Here, $\mathbf{G}$ and $\mathbf{D}$ are discrete approximations
 of the SRT operator (Eqn.~(\ref{eq:2b_SRT})) and the differential operator
 (Eqn.~(\ref{eq:2b_g_def})), respectively. 

The discrete SRT operator $\mathbf{G}$ can be implemented in a ``temporal-sample-driven" manner;
namely, the pressure data are computed by accumulating the contributions from voxels on a discretized spherical shell surface defined by the current data sample \cite{wang2013accelerating}: 
\begin{equation}
\label{eq:SRTd}
	 [\mathbf{G} \boldsymbol{\theta}]_{qK+k} \equiv \Delta_s^2 \sum_{n=0}^{N-1} [\boldsymbol{\theta}]_n \sum_{i=0}^{N_i-1} \sum_{j=0}^{N_j-1} \phi_n^{int}(\mathbf{r}_{k,i,j}) \equiv [\mathbf{g}]_{qK+k},
\end{equation}
where $N_i$, $N_j$ denotes the numbers of angular divisions over the polar and azimuth directions within the local spherical coordinate system centered at the $q$-th transducer $\mathbf{r}_q'$ with a radius of $k \Delta_t c_0$, in which the center of $i$-th polar division and $j$-th azimuth division is denoted by $\mathbf{r}_{k, i, j}$ ~\cite{wang2013accelerating}. 

The differential operator $\mathbf{D}$ can be implemented as 
\begin{equation}
\label{eq:differentiald}
	[\mathbf{Dg}]_{qK+k} \equiv \frac{1}{8 \pi c_0^2 \Delta_t^2} \left( \frac{[\mathbf{g}]_{qK+k+1}}{k+1} - \frac{[\mathbf{g}]_{qK+k-1}}{k-1} \right).
\end{equation}
Due to their large sizes,  the explicit storage of system matrices is typically infeasible with current computer technologies except for small scale problems.
By use of Eqns.\ (\ref{eq:SRTd}) and (\ref{eq:differentiald}) the action of
$ \mathbf{H}_{int}$ can be computed without having to explicitly store its elements.

\subsubsection{Kaiser-Bessel function-based D-D PACT model}
\label{subsec:3b_2}
The Kaiser-Bessel (KB) function-based D-D PACT model employs the KB functions of order $m$ as the expansion functions. These KB functions are defined as
 \cite{lewitt1990multidimensional,schweiger2017basis,wang2014discrete}
\begin{equation}
	b(x) = \left\{ \begin{array}{lc} 
			\left( \sqrt{1-x^2/a^2} \right)^m \frac{I_m(\gamma \sqrt{1-x^2/a^2})}{I_m(\gamma)}, & \text{if } 0 \leq x \leq a	\\
			0, & \text{if } x > a, 
		\end{array} \right.
\end{equation}
where $x \in \mathbb{R}^+$, $a \in \mathbb{R}^+$ denotes the support radius of the KB function, $\gamma \in \mathbb{R}^+$ describes the smoothness of the KB function. 
The term $I_m(x)$ denotes the modified Bessel function of the first kind of order $m$. 
The associated expansion function can be expressed as 
\begin{equation}
	\phi^{\rm KB}_n(\mathbf{r}) = b(|\mathbf{r} - \mathbf{r}_n|),
\end{equation}
which is simply the KB function centered at location $\mathbf{r}_n$. 

When employing the KB expansion functions, it is convenient to formulate the corresponding D-D PACT imaging model in the temporal frequency domain \cite{wang2012investigation}. 
Let $\tilde{\mathbf{u}}$ denotes the discrete Fourier transform (DFT)
 of the measurement data $\mathbf{u}$. 
The KB function-based D-D PACT imaging model can be expressed as \cite{wang2014discrete}
\begin{equation}
	\tilde{\mathbf{u}} = \tilde{\mathbf{H}}_{KB} \boldsymbol{\theta}_{KB},
\end{equation}
where the elements of the system matrix $\tilde{\mathbf{H}}_{KB}$ can be written as the multiplication of two terms: 
\begin{gather}
	[\tilde{\mathbf{H}}_{KB}]_{qL+l, n} \equiv \tilde{p}^{KB}(f) \tilde{h}_q^{s, point}(\mathbf{r}_n, f) \vert_{f = l\Delta_f}, \text{ for } l = 0, 1, \dots, L-1,
	\label{eq:3b_HKB}
\end{gather}
where 
\begin{gather}
	\tilde{p}^{KB}(f) = - \frac{j 4\pi^2 f a^3 \beta}{C_p I_m(\gamma)} \frac{\hat{j}_{m+1}(\sqrt{4 \pi^2 a^2 f^2 / c_0^2 - \gamma^2})} {(4\pi^2 a^2 f^2 / c_0^2 - \gamma^2)^{(m+1)/2}}, 
	\nonumber 
\end{gather}
and
\begin{gather}
	\tilde{h}_q^{s, point}(\mathbf{r}_n, f) = \frac{\exp\left(-j 2\pi f \frac{|\mathbf{r}_q^s - \mathbf{r}_n|}{c_0}\right)}{2 \pi |\mathbf{r}_q^s - \mathbf{r}_n|}.
	\label{eq:3b_SIR_Green}
\end{gather}
Here, $L$ denotes the total number of frequency components and $\Delta_f$ is the frequency sampling interval. 
The first quantity $\tilde{p}^{KB}(f)$ in Eqn.~(\ref{eq:3b_HKB}) is the temporal Fourier transform of the acoustic pressure generated by a KB function located at origin, where $\hat{j}_m(x)$ is the $m$-th order spherical Bessel function of the first kind \cite{diebold2009photoacoustic}. 
The second quantity $\tilde{h}_q^{s, point}(\mathbf{r}_n ,f)$ is the spatial impulse response (SIR) in the temporal frequency domain. 
Under the assumption of an idealized point-transducer model, this quantity reduces to the Green function given in Eqn.~(\ref{eq:3b_SIR_Green}), where $\mathbf{r}_q^s$ denotes the location of the $q$-th transducer and $\mathbf{r}_n$ denotes the location of the center of the $n$-th KB expansion function \cite{wang2014discrete}.

\subsection{D-D forward models for use with heterogeneous acoustic media}
\label{subsec:section3c}

\subsubsection{Overview of approaches}

The establishment of  D-D imaging models for lossy, heterogeneous fluid media requires the introduction of finite-dimensional approximations of the object function $p_0(\mathbf{r})$ as well as the known medium acoustic parameters such
 as $c(\mathbf{r}), \rho_0(\mathbf{r})$ and $\alpha(\mathbf{r})$. In principle, these medium acoustic parameters can be estimated
from adjunct ultrasound tomography data \cite{Manohar:concomitant,Jinxing:acoustic}.  
When such adjunct image data are not available, one can attempt 
to estimate the acoustic parameters concurrently with $p_0(\mathbf r)$
as described in Section~\ref{sec:section5}.
 For PACT in heterogeneous media, the choice of finite-dimensional representation of the object and medium parameters is dictated by the numerical method employed to solve the coupled first order differential equations described in Eqn.~\eref{eq:2a_loss}.
 Most methods for computing a numerical solution of
 the acoustic wave equation in lossy heterogeneous media fall into one of three major categories: finite difference (FD) methods, finite element (FE) methods, and spectral methods.

The FD and FE methods are also known as local methods because the wave propagation equations are solved at each point based only on conditions at neighboring points. In contrast,
 spectral methods are global,
 as information from the entire wavefield is employed
 to numerically propagate the wavefield.
 As the spectral methods leverage global information,
 they allow computation to be performed on coarser grid while maintaining
 high accuracy~\cite{gottlieb1977numerical}. As opposed to FD or FE methods that require grid spacing on the order 10 points per minimum wavelength,  spectral methods, in theory, only require 2 points per minimum wavelength~\cite{cox2007k}.
 Due to such relaxed spatial sampling constraints, spectral methods are well suited for large-scale, high frequency PA wave solvers. One of the most popular spectral methods used for solving the acoustic wave equation is the k-space pseudospectral method~\cite{mast2001k,tabei2002k,cox2007k}.
Because of its popularity in the PACT community,  the D-D imaging model
presented below will be based upon the k-space pseudospectral time-domain (PSTD) method \cite{cox2007k}.

\subsubsection{D-D forward model based on the k-space PSTD method }
\label{sec:DDPS}
Below, a general formulation of a D-D  forward model
based on the k-space PSTD method is briefly outlined.  Because of
the highly technical nature of the formulation, the reader
is referred to the literature for  specific details \cite{huang2013full}.

 The finite-dimensional approximations of the object function $p_0(\mathbf{r})$ as well as the medium parameters $c(\mathbf{r}), \rho_0(\mathbf{r})$ and $\alpha(\mathbf{r})$ are constructed by sampling these quantities on a Cartesian grid. If $\{\phi_n^{\rm int}(\mathbf{r})\}_{n = 0}^{N-1}$ denotes the set of 3D Cartesian expansion functions defined in Eqn.~\eref{eq:3b_interpolation_expansion}, the initial pressure distribution $p_0(\mathbf{r})$ can be approximated as a finite dimensional vector $\boldsymbol{\theta} \in \mathbb{R}^{N \times 1}$.
To ensure the stability of the k-space PSTD method, the material properties such as the speed of sound $c(\mathbf{r})$, the ambient density $\rho_0(\mathbf{r})$, the absorption coefficient $\alpha(\mathbf{r})$, and the wavefield parameters such as pressure $p(\mathbf{r}, t)$,  particle velocity $\dot{\mathbf{s}}(\mathbf{r}, t)$, and acoustic density $\rho(\mathbf{r},t)$,
 are sampled at different points of staggered Cartesian grids~\cite{treeby2010k}. Furthermore, the wavefield parameters such as the pressure, particle velocity and acoustic density are staggered temporally. Thus, the finite-dimensional representations of the medium parameters and the wavefield parameters for use in a D-D imaging model will generally employ different expansion functions.

Let $\dot{\mathbf{s}}^i_k \in \mathbb{R}^{N \times 1}$, $\boldsymbol{\rho}^i_k \in \mathbb{R}^{N \times 1}$ represent the discrete approximations of the vector-valued particle velocity and acoustic density over the whole 3D Cartesian grid at the $k^{th}$
 time step along the $i^{th}$ direction. Let $\mathbf{p}_k \in \mathbb{R}^{N \times 1}$ represent the acoustic pressure sampled at all spatial grid points
 at the $k^{th}$ time step.

 Let $\mathbf{v}_k = (\dot{\mathbf{s}}_k^1, \dot{\mathbf{s}}_k^2, \dot{\mathbf{s}}_k^3, \boldsymbol{\rho}_k^1, \boldsymbol{\rho}_k^2, \boldsymbol{\rho}_k^3, \mathbf{p}_k)^T$ denote a $7N \times 1$ vector containing all the wavefield variables at the time step $k\Delta t$.  The image acquisition process in PACT can be mathematically modeled as the propagation of the wavefield parameters forward in time from $t = 0$ to $t = (K-1)\Delta t$ as 
\begin{align}\label{eq:3c_Prop}
\begin{bmatrix}\mathbf{v}_0, \mathbf{v}_1, \cdots, \mathbf{v}_{K-1}
\end{bmatrix}^T = \mathbf{T}_{K-1} \cdot \mathbf{T}_{K-2} \cdots \mathbf{T}_1
\begin{bmatrix}
\mathbf{v}_0,\mathbf{0}_{7N \times 1},\cdots,\mathbf{0}_{7N \times 1}
\end{bmatrix}^T.
\end{align}
For specific details regarding the definition of the temporal matrix $\mathbf{T}_k (k = 1,\cdots, K-1) \in \mathbb{R}^{7NK \times 7NK} $, the reader
is referred to the literature~\cite{huang2013full}.
From the initial conditions defined in Eqn.~\eref{eq:2a_loss}, the vector $(\mathbf{v}_0, \mathbf{0}_{7N\times 1}, \cdots, \mathbf{0}_{7N \times 1})^T$ can be computed from the initial pressure distribution $\boldsymbol{\theta}$ as 
\begin{align}\label{eq:3c_Init}
\begin{bmatrix}
\mathbf{v}_0, \mathbf{0}_{7N \times 1}, \cdots, \mathbf{0}_{7N \times 1}
\end{bmatrix}^T = \mathbf{T}_0 \boldsymbol{\theta},
\end{align}
where $\mathbf{T}_0 \in \mathbb{R}^{7N \times N}$ maps the initial pressure distribution to the initial value of the wavefield variables at time $t = 0$~\cite{huang2013full}.

In general, the transducer locations $\mathbf{r}_q^{\prime}$ at which the pressure are recorded will not coincide with the vertices of the 3D Cartesian grid at which the propagated wavefields are computed. The pressure at the transducer locations can be related to the computed field quantities via an interpolation operation defined as 
\begin{align}\label{eq:3c_Interp}
\mathbf{u} = \mathbf{M}\begin{bmatrix}
\mathbf{v}_0, \mathbf{v}_1, \cdots, \mathbf{v}_{K-1}
\end{bmatrix}^T,
\end{align}
where $\mathbf{M} \in \mathbb{R}^{QK \times 7NM}$. The choice of the interpolation operation is a parameter that will guide the construction of the matrix $\mathbf{M}$. Some of the most commonly employed interpolation strategies include trilinear interpolation or Delaunay triangulation-based interpolation~\cite{lee1980two}. 

By use of Eqns.~\eref{eq:3c_Init},~\eref{eq:3c_Prop} and~\eref{eq:3c_Interp}, one obtains
\begin{align}\label{eq:3c_Imageeqn}
\mathbf{u} = \mathbf{M}\mathbf{T}_{K-1}\cdots \mathbf{T}_1 \mathbf{T}_0 \boldsymbol{\theta}.
\end{align}
The explicit form of the system matrix $\mathbf{H}_{PS} \in \mathbb{R}^{QK \times N}$, that describes the propagation of PA waves based on Eqn.~\eref{eq:2a_loss} can be therefore expressed as
\begin{align}\label{eq:3c_HPS}
\mathbf{H}_{PS} = \mathbf{M}\mathbf{T}_{K-1}\cdots \mathbf{T}_1 \mathbf{T}_0.
\end{align}
Here, the subscript $PS$ stands for the fact that the D-D imaging model is
 computed using the k-space PSTD method.

\subsection{Incorporation of transducer responses in D-D forward models}
\label{sec:3d}

In practice, the ultrasound transducers employed in a PACT imager are imperfect and result in measurements of the PA wavefield that are averaged over finite temporal and spatial apertures as described in Eqn.\ (\ref{eq:CDaperture}).
In the ultrasound imaging community, the effects of these sampling apertures are characterized by the transducer's electrical impulse response (EIR) and
 spatial impulse response (SIR).
Failure to account for these effects can result in reconstructed estimates of $p_0(\mathbf r)$ that contain distortions and/or degraded spatial resolution \cite{xu2003analytic}.


When iterative image reconstruction methods are to-be-employed,  a natural
 way to compensate for the EIR and SIR is to include them in the constructed D-D forward model
\cite{wang2011imaging,rosenthal2011model,ding2017efficient}.
Below,  the D-D forward models for use with homogeneous acoustic media introduced in Section~\ref{subsec:3b}
are extended to incorporate the transducer responses.
The extension of the the D-D forward model for use with heterogeneous
 acoustic media introduced in Section~\ref{subsec:section3c} to include transducer responses is relatively straightforward
and is described in the literature \cite{huang2013full}.


\subsubsection{Incorporating the EIR in D-D forward models for use with homogeneous acoustic media}
\label{subsubsec:3d_ii}

Because degradation of the measured data by the EIR is described by a linear time-invariant model, it can be readily incorporated into the system matrices.
For example, the interpolation-based system matrix with EIR can be re-defined as \cite{wang2013accelerating}
\begin{equation}
    \mathbf{H}_{int} = \mathbf{H}^e \mathbf{DG},
    \label{eq:3d_int_DD}
\end{equation}
where $\mathbf{D}, \mathbf{G}$ are the discrete approximation so the differential operator and the SRT operator and  
\begin{equation}
    \mathbf{H}^e \mathbf{p}_{ideal} \equiv \mathcal{F}^{-1}\{ \mathcal{F}(\mathbf{h}^e) \mathcal{F}(\mathbf{p}_{ideal}) \}.
\end{equation}
Here, $[\mathbf{h}^e]_k = \Delta_t h^e(t) \vert_{t = k \Delta t}$ is the EIR signal sampled in the
 temporal domain, $\mathcal{F}$ and $\mathcal{F}^{-1}$ represent the discrete Fourier transform and inverse discrete Fourier transform, respectively, and $\mathbf{p}_{ideal}$ denotes the temporally sampled ideal pressure
 signal vector that would be recorded by an idealized point transducer.

A similar approach can be adopted to incorporate EIR in the KB function-based system matrix.
In this case,  the convolution operation can be implemented as an element-wise multiplication in the temporal frequency domain \cite{wang2012investigation,wang2013accelerating} and the elements of the system matrix
are re-defined as
\begin{equation}
    [\mathbf{H}_{KB}]_{qL+l, n} = \tilde{p}^{KB}(f) \tilde{h}^e(f), \vert_{f = l \Delta_f},\text{ for } l = 0, 1, \dots, L-1,
\end{equation}
where samples of  $\tilde{h}^e(f)$ are computed as the discrete
 Fourier transform of $\mathbf{h}^e$.

\subsubsection{Incorporating the SIR in D-D forward models for use with homogeneous acoustic media}
The spatial impulse response (SIR) describes the spatial averaging of an
 acoustic signal that occurs when the signal is measured by use of a transducer possessing
a non-zero active area.  Specifically, consider a  point acoustic source located
 at position $\mathbf{r}_n$ whose temporal response is described by $\delta(t)$.
 The SIR ${h}_q^s(\mathbf{r}_n,t)$
represents the measurement of the radiated wavefield
  by a transducer indexed by $q$ that has an idealized EIR.

Various SIR models have been proposed  \cite{harris1981review,lockwood1973high,stepanishen1971transient} and
employed in PACT \cite{ermilov2009development,wang2011imaging,wang2012investigation,mitsuhashi2014investigation,rosenthal2011model,queiros2013modeling,ding2017efficient,wiskin2012non,sandhu2015frequency}. Because the degradation of the measured signal by the SIR is not generally described by a linear time-invariant model, it is not as straightforward to compensate for as is the EIR.

For the interpolation-based D-D imaging model in Section~\ref{subsec:3b}, Ding et al.\
 proposed to approximately incorporate the SIR 
by summing the pressure signal over a collection of points on the transducer
 surface. 
However, in order to accurately model the effects of the SIR,
it may be necessary to sample a large number of
 points on the transducer surface, which can result in a large computational burden.
It remains generally difficult to accurately incorporate the SIR
in an interpolation-based PACT D-D forward model and therefore certain
 implementations of this model have ignored the SIR \cite{wang2013accelerating}. 

In the KB function-based system matrix, the SIR can be readily incorporated as an additional element-wise multiplication step in the temporal frequency domain:
\begin{equation}\label{eq:3d_ii_forward}
    [\mathbf{H}_{KB}]_{qL+l, n} = \tilde{p}^{KB}(f) \tilde{h}^e(f) \tilde{h}_q^s(\mathbf{r}_n, f) \vert_{f = l \Delta_f}.
\end{equation}
Here, $\tilde{h}_q^s(\mathbf{r}_n,f)$ denotes the temporal Fourier transform of ${h}_q^s(\mathbf{r}_n,t)$ that can be computed by
 integrating the Green function in Eqn.~(\ref{eq:3b_SIR_Green}) over the $q$-th transducer surface $S_q$:
\begin{equation}
    \tilde{h}_q^s(\mathbf{r}_n, f) = \int_{S_q} \diff \mathbf{r}^3 \frac{\exp (-j 2\pi f \frac{|\mathbf{r}_q^s - \mathbf{r}_n|}{c_0} )}{2 \pi |\mathbf{r}_q^s - \mathbf{r}_n|}.
    \label{eq:3d_SIR_1}
\end{equation}
Note that the integral in Eqn.~\eref{eq:3d_SIR_1} resembles the Rayleigh integral~\cite{kirkup1994computational}.
As a specific example, consider a transducer element that possesses a
  rectangular detecting surface of area $a \times b$.
Under the validity of the far field assumption \cite{mitsuhashi2014investigation,born2013principles}:
\begin{equation}
    |\mathbf{r}_q^s - \mathbf{r}_n| \gg \frac{\max(a,b)^2}{\lambda},
    \label{eq:3d_far_field}
\end{equation}
Eqn.~(\ref{eq:3d_SIR_1}) can be evaluated to determine
  $\tilde{h}_q^s(\mathbf{r}_n, f)$  as
 \cite{stepanishen1971transient}
\begin{equation}
    \tilde{h}_q^s(\mathbf{r}_n, f) = \frac{ab \exp(-j 2\pi f \frac{|\mathbf{r}^s_q - \mathbf{r}_n|}{c_0})}{2\pi |\mathbf{r}^s_q - \mathbf{r}_n|} \sinc \left( \pi f \frac{a x_{nq}^{tr}}{c_0 |\mathbf{r}_q^s - \mathbf{r}_n|} \right) \sinc \left( \pi f \frac{b y_{nq}^{tr}}{c_0 |\mathbf{r}_q^s - \mathbf{r}_n|} \right),
    \label{eq:3d_SIR_2}
\end{equation}
where $x_{nq}^{tr}, y_{nq}^{tr}$ are the transverse components of $\mathbf{r}_n$
 in a local coordinate system centered at $\mathbf{r}_q^s$.
 Given $\mathbf{r}_n = (x_n, y_n, z_n)$ and the transducer location $\mathbf{r}^s_q$ specified in spherical polar coordinates $\mathbf{r}^s_q = (r_q, \theta_q, \phi_q)$, the values of the transverse coordinates can be computed as \cite{wang2012investigation}:
\begin{gather}
    x_{nq}^{tr} = -x_n \cos \theta_q \cos \phi_q - y_n \cos \theta_q \sin \phi_q + z_n \sin \theta_q,       \nonumber \\
    y_{nq}^{tr} = -x_n \sin \phi_q + y_n \cos \phi_q.
\end{gather}

\subsubsection{Patch-based estimation of the SIR}

When the far field condition in Eqn.\ (\ref{eq:3d_far_field}) is violated,
 the SIR expression in Eqn.~(\ref{eq:3d_SIR_2}) can be inaccurate and its use
 may lead to  artifacts in the reconstructed images \cite{mitsuhashi2014investigation}.
Moreover, when the transducer surface is not flat, it may be difficult to determine
an analytic expression for the SIR.
These issues can be mitigated by use of ``divide-and-integrate" approaches, including line-detector-based method \cite{rosenthal2011model} and patch-based method \cite{mitsuhashi2014investigation}. In such approaches, the surface of the transducer is computationally decomposed
 into smaller elements whose SIRs can be more readily determined.
In the line-detector based model, each transducer element surface is decomposed into a number of parallel straight lines, whose SIRs can be analytically expressed.
In the patch-based method, each transducer element surface is divided into smaller planar patches that each satisfy the far field
 approximation \cite{mitsuhashi2014investigation}; 
 subsequently, in either case, the SIR of the transducer can be approximated by computing the average of the SIRs of the sub-elements (lines or patches):
\begin{equation}
    \tilde{h}^s_q(\mathbf{r}_n, f) \approx \frac{1}{N_{patch}} \sum_{i=1}^{N_{patch}} \tilde{h}^s_{q, i}(\mathbf{r}_n, f),
    \label{eq:3d_SIR_patch}
\end{equation}
where $\tilde{h}_{q,i}^s(\mathbf{r}_n, f)$ denotes the SIR corresponding to  $i$-th sub-element.
In the patch-based approach, $\tilde{h}_{q,i}^s(\mathbf{r}_n, f)$ can be computed with the aid of Eqn.~(\ref{eq:3d_SIR_2}).
It should be noted that, in the line-detector model, the approximation only accounts for the SIR effect in the direction that is parallel to the straight lines, it still assumes zero thickness (point-like) transducer in the perpendicular direction. 
The patch-based model employs the far-field approximation for both directions in the transducer plane, therefore providing compensation in both directions.
In addition, the patch-based model can be extended beyond planar transducers by decomposing an arbitrary transducer surface into smaller patches to estimate its SIR.

\begin{figure}[htbp]
    \centering
    \includegraphics[width = 1.0 \linewidth]{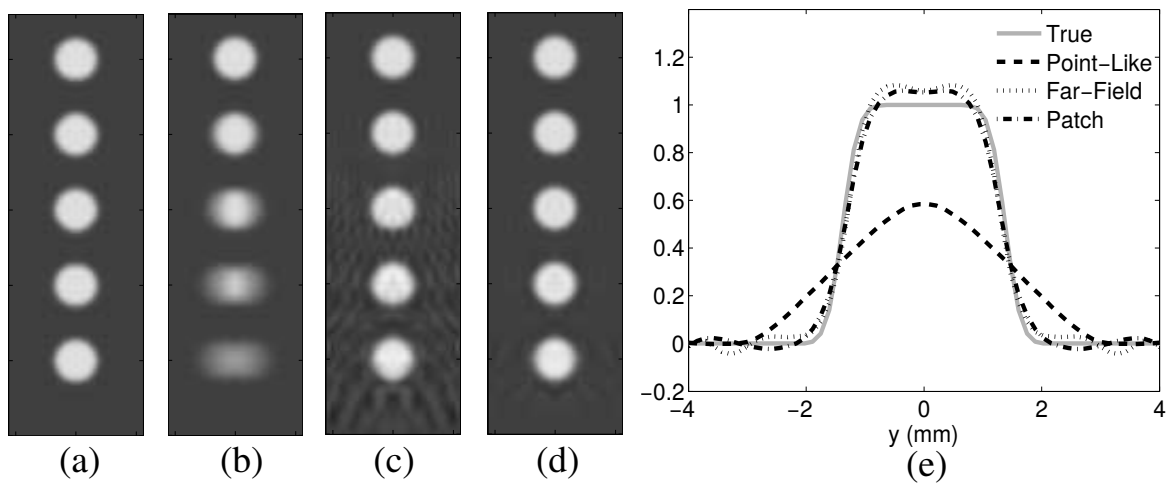}
    \caption{Examples showing the effect of different SIR models in PACT reconstruction algorithms.
        The lowest sphere is closest to the detector element.
        (a) The center Z slice of the original phantom.
        (b) Image reconstructed assuming point-like transducer models.
        (c) Image reconstructed assuming far-field-based SIR model.
        (d) Image reconstructed assuming patch-based SIR model.
        (e) The line profiles through the center of the lower-most sphere. The images were reproduced from the literature~\protect\cite{mitsuhashi2014investigation}.}
    \label{fig:SIR_example}
\end{figure}

A computer simulation study examining the effects of accurately modeling the SIR for flat rectangular transducers in the context of PACT image reconstruction is discussed below~\cite{mitsuhashi2014investigation}. The numerical phantom used for the study consisted of spherical objects placed within a PACT system as shown in Fig.~\ref{fig:SIR_example}(a).
The simulated pressure data for each sphere were generated by numerically convolving a closed form expression of waves generated by a solid sphere \cite{diebold2009photoacoustic}, with a semi-analytical SIR specifically for spherical waves \cite{jensen1999new}. 
During reconstruction, the Kaiser-Bessel function-based D-D PACT imaging model introduced in Section \ref{subsec:3b_2} was employed with different SIR models: 
    Fig.~\ref{fig:SIR_example}(b) assumes a point-like transducer model, 
    Fig.~\ref{fig:SIR_example}(c) employs the far-field-based SIR model as in Eqns.~(\ref{eq:3d_SIR_1}) and (\ref{eq:3d_SIR_2}), 
    and Fig.~\ref{fig:SIR_example}(d) employs the patch-based model in Eqn.~(\ref{eq:3d_SIR_patch}).
The reconstructed images where obtained by solving a penalized-least-squares optimization problem with quadratic smoothness penalty using the conjugate gradient method. 
These results show that ignoring transducer SIR in the PACT imaging model leads to significant distortion in reconstructed images. 
Incorporating an accurate transducer SIR model can greatly reduce such artifacts. For objects far away from the transducers, the far-field-based SIR model may suffice. However, for closer objects, patch-based or other "divide-and-integrate" SIR models can further improve the reconstruction accuracy.

\section{Image reconstruction approaches}
\label{sec:section4}

In this section, a brief survey of conventional image reconstruction methods for PACT is provided.
Subsequently, optimization-based image reconstruction methods that are
based upon D-D forward models are presented in Section~\ref{subsec:4c}.

\subsection{Brief overview of analytic reconstruction methods}
\label{subsec:4a}
 A large number of analytic, or non-iterative, tomographic reconstruction methods for PACT are
 currently available. Several review articles provide detailed
descriptions of these \cite{kuchment2011mathematics,Li-Review}, so only a quick overview is provided here.
Analytic reconstruction methods are commonly based upon a C-C PACT forward model that assumes
a homogeneous acoustic media, such as those described in Section~\ref{subsec:2b}.
Analytic reconstruction formulae of filtered backprojection (FBP) form are available for special detector
 geometries~\cite{finch2004determining,Xu:2005bp,kunyansky2007explicit,finch2007inversion,kunyansky2011reconstruction,haltmeier2014universal}.
Such reconstruction methods typically assume an acoustically homogeneous and lossless medium
and full-view acoustic detection in which the pressure measurements are densely sampled on a closed surface
that encloses the object.  Additionally, the available analytic methods are  unable to compensate for the
ultrasound transducer responses~\cite{wang2011imaging}.
Other ad hoc reconstruction techniques such as inversion of the Radon transform~\cite{Kruger:95}, as well as delay-and-sum techniques~\cite{hoelen2000image} have been employed for PACT image reconstruction.

PACT reconstruction methods that are based on harmonic decomposition  have
 also been proposed~\cite{Norton:81}.
 For special geometries such as planar detector geometries, such reconstruction  methods
 can be implemented efficiently by use of the fast Fourier transform (FFT)~\cite{kostli,fawcett1985inversion}.
 This approach has been extended for use with novel measurement geometries that
utilize reverberant cavities~\cite{cox2007photoacoustic}, and
 for use with closed measurement surfaces for which the eigenfunction
 of the Dirichlet Laplacian are explicitly known~\cite{kunyansky2007series}.
 These approaches possess computationally efficient implementations and, in certain cases,
can reconstruct 3D images thousands of times faster than
 backprojection-type methods~\cite{kunyansky2007series}.
 Similar to the FBP-type algorithms, the harmonic decomposition-based reconstruction methods
typically assume an acoustically homogeneous media and do not compensate for the transducer responses.



\subsection{Time-reversal reconstruction methods}
\label{subsec:4b}

As an alternative to the analytic reconstruction approaches discussed above,
a variety of
  time-reversal (TR) based reconstruction methods have
 been proposed~\cite{xu2004time,stefanov2009thermoacoustic,hristova2008reconstruction,treeby2010photoacoustic,palacios2016reconstruction}.
 Image reconstruction via TR is achieved by running a numerical model of the wave equation-based forward problem
 backwards in time. Namely,
 the measured acoustic pressure signals are re-transmitted into the domain in a time-reversed fashion. 
As such, they can readily account for wave propagation in heterogeneous media and
can accommodate arbitrary detection geometries.

\subsubsection{Formulation of TR image reconstruction for PACT}

Consider a compactly supported initial pressure distribution $p_0(\mathbf r)$ in an infinite 3D homogeneous lossless acoustic medium,
and let the domain $\Omega$ correspond to the interior of a closed measurement surface $\Omega_0$ that
encloses the to-be-imaged object.
According to Huygen's principle, 
there exists a time $T>0$ for which the radiated wavefield inside $\Omega$ vanishes  $\forall t\geq T$.
 Because $\Omega$ contains no energy  $\forall t\geq T$,
 one can  solve the wave equation backwards in time inside $\Omega$ with zero initial conditions and the boundary condition given by the measured data on the surface $\Omega_0$.  As described
mathematically below, this process of rewinding the waves backward in time to reconstruct the $p_0(\mathbf r)$ is defined as TR. 
When finite sampling effects are ignored, it has been demonstrated that time reversal yields a theoretically exact reconstruction of $p_0(\mathbf r)$ for the case of a 3D acoustically homogeneous medium 
 if $T$ is large enough to allow the energy to escape the domain $\Omega$~\cite{agranovsky2007uniqueness}. 
In even dimensions or when the medium is acoustically heterogeneous, this result does not hold true.
 Nevertheless, the TR method can still be employed to obtain an estimate of $p_0(\mathbf r)$.
When the speed of sound distribution is non-trapping (i.e. all of the acoustic energy escapes the domain $\Omega$),
 the errors in the estimates can be bounded as a function of the acquisition time $T$~\cite{hristova2009time}.

In addition, the same principle of replaying the wavefield back in time can also be applied in lossy fluid media. Intuitively, to compensate for the amplitude loss in the wavefields, the wavefields should be corrected by
 gain factors. To account for acoustic absorption, the absorption term in Eqn.~\eref{eq:2a_lossiii} must be reversed in sign when computing the time-reversed wavefields. In contrast, to account for the dispersion, the sign of the dispersive speed is left unchanged~\cite{treeby2010photoacoustic}. Thus, the TR reconstruction
method for use with  lossy heterogeneous media solves the following system of equations for the
time-reversed pressure wavefield $p(\mathbf r, t)$ ~\cite{treeby2010photoacoustic}:
\begin{subequations}\label{eq:4b_lossTR}
	\begin{align}
	\frac{\partial}{\partial t} \dot{\mathbf{s}}(\mathbf{r},t) &= - \frac{1}{\rho_0(\mathbf{r},t)} \nabla p(\mathbf{r},t), \label{eq:4b_lossTRi}\\
	\frac{\partial}{\partial t} \rho(\mathbf{r},t) &=  - \rho_0(\mathbf{r}) \nabla \cdot \dot{\mathbf{s}}(\mathbf{r},t), \label{eq:4b_lossTRii}\\
	p(\mathbf{r},t) 
	= c_0(\mathbf{r})^2 &\Big\{ 1 + \mu(\mathbf{r}) \frac{\partial}{\partial t} (-\nabla^2)^{\frac{y}{2} - 1} - \eta(\mathbf{r}) (-\nabla^2)^{\frac{y-1}{2}}\Big\} \rho(\mathbf{r},t),\label{eq:4b_lossTRiii} \\
	\textnormal{subject to the conditions: }&\nonumber \\
	p(\mathbf{r},t)\Big|_{t = 0} = 0,\
	\ &\dot{\mathbf{s}}(\mathbf{r},t)\Big|_{t= 0}= 0 \textnormal{ and }
	p(\mathbf{r},t)\Big|_{\mathbf{r} \in \Omega_0} = u(\mathbf{r},T -t)\Big|_{\mathbf{r} \in \Omega_0}\ .
	\end{align}
\end{subequations}
Here, $u(\mathbf{r}, t)$ denotes the measured pressure data for the case
where finite sampling effects are neglected and idealized point-like ultrasound transducer are employed.
The sought after estimate of $p_0(\mathbf r)$ is specified by the TR method as
\begin{align}\label{eq:4b_TRfin}
\hat{p}_0^{TR}(\mathbf{r}) \equiv p(\mathbf{r},t= T) \textnormal{, } \forall \textnormal{ }\mathbf{r} \in \Omega.
\end{align}
In operator form,  Eqns.~\eref{eq:4b_lossTR} and~\eref{eq:4b_TRfin} can be expressed as
\begin{align}
\hat{p}_0^{TR}(\mathbf{r}) = \mathcal{H}^{TR}u(\mathbf{r}, t),
\end{align}
where $\mathcal{H}^{TR}:\mathbb{L}_2(\Omega_0 \times [0,T]) \mapsto \mathbb{L}_2(\Omega)$ denotes the time reversal operator described in Eqn.~\eref{eq:4b_lossTR}.

\subsubsection{Modified TR image reconstruction based on a Neumann series method}
As described above, the canonical TR method
 is mathematically exact when the radiated wavefield $p(\mathbf{r},t)$
 decays to zero inside the domain $\Omega$ for some finite time $T$.
 However, in even dimensions and/or when the medium is acoustically
 heterogeneous and non-trapping, this condition may not be satisfied~\cite{hristova2008reconstruction,hristova2009time}.
A reconstruction method based on a Neumann series (NS) expansion has been developed that can be employed
for accurate image reconstruction
in the presence of acoustically heterogeneous media and an irregular observation geometry
 for a finite acquisition time $T$ \cite{qian2011efficient,stefanov2009thermoacoustic}.
 The NS-based reconstruction method is accurate when the pressure $p(\mathbf{r},t)$ is known on the whole boundary $\Omega_0$ and $T$ is greater than a stability threshold~\cite{qian2011efficient}.
 To define the NS-based reconstruction method, one needs to first introduce the modified TR operator. The modified TR operator for lossy heterogeneous media solves Eqns.~\eref{eq:4b_lossTRi}, \eref{eq:4b_lossTRii} and \eref{eq:4b_lossTRiii} subject to the initial and boundary value conditions:  
\begin{align}\label{eq:4b_modTR}
	p(\mathbf{r},t)\Big|_{t = 0} &= P_\Omega u(\mathbf{r},T),\
	\ \dot{\mathbf{s}}(\mathbf{r},t)\Big|_{t= 0}= 0 \textnormal{ and }
	p(\mathbf{r},t)\Big|_{\mathbf{r} \in \Omega_0} = u(\mathbf{r},T -t)\Big|_{\mathbf{r} \in \Omega_0}\ ,
\end{align}
where $P_\Omega: \mathbb{L}_2(\Omega_0) \mapsto \mathbb{L}_2(\Omega)$ is a harmonic extension operator. The harmonic extension operator is defined as $P_\Omega g(\mathbf{r}, T) = \phi(\mathbf{r})$, where $\phi(\mathbf{r})$ is the solution to the elliptic boundary value problem 
\begin{align}\label{eq:4b_Pos}
\Delta \phi(\mathbf{r}) = 0, \ \ \phi(\mathbf{r})\Big|_{\mathbf{r} \in \Omega_0} = g(\mathbf{r}, T)\Big|_{\mathbf{r} \in \Omega_0}.
\end{align}
One can summarize Eqns.~\eref{eq:4b_modTR}, ~\eref{eq:4b_TRfin} and~\eref{eq:4b_Pos} as
\begin{align}
\hat{p}_0^{TR}(\mathbf{r}) = \mathcal{H}_{mod}^{TR} u(\mathbf{r}, t),
\end{align}
where $\mathcal{H}_{mod}^{TR}:\mathbb{L}_2(\Omega_0 \times [0,T]) \mapsto \mathbb{L}_2(\Omega)$ is the
 modified TR reconstruction operator specified by Eqns.~\eref{eq:4b_modTR} and~\eref{eq:4b_Pos}.

Given the modified TR operator, the NS-based reconstructed initial pressure distribution can be defined as 
\begin{align}\label{eq:4b_NS}
\hat{p}_0^{NS}(\mathbf{r}) = \sum_{m = 0}^{\infty} (I - \mathcal{H}_{mod}^{TR} \mathcal{H}_{wave})^m \mathcal{H}_{mod}^{TR}u(\mathbf{r},t),
\end{align}
where $\mathcal{H}_{wave}$ is the C-C wave equation-based forward operator in Eqn.\ (\ref{eq:2a_HCC}) and
 $I$ is the identity operator.  Although the NS-based reconstruction is guaranteed to be convergent when the speed of sound distribution is non-trapping, it has been successfully applied to the non-trapping case as well~\cite{qian2011efficient}.

As the sum defined in Eqn.~\eref{eq:4b_NS} extends from $m = 0$ to $m = \infty$, it cannot exactly be implemented. It is interesting to note that the NS method can be interpreted as an iterative time reversal method~\cite{arridge2016adjoint}. Let us define the $i^{th}$ estimate of the reconstructed initial pressure distribution as
\begin{align}\label{eq:4b_TrunkNS}
	\hat{p}_0^{NS,i}(\mathbf{r}) = \sum_{m = 0}^{i}(I - \mathcal{H}_{mod}^{TR} \mathcal{H}_{wave})^m \mathcal{H}_{mod}^{TR}u(\mathbf{r},t).
\end{align}
The partial sum in Eqn.~\eref{eq:4b_TrunkNS} can be expressed as  
\begin{align}\label{eq:4b_Nsit}
	\hat{p}_0^{NS,i}(\mathbf{r}) &= \hat{p}_0^{NS,i -1}(\mathbf{r})  - \mathcal{H}_{mod}^{TR}(\mathcal{H}_{wave} \hat{p}_0^{NS,i-1}(\mathbf{r}) - \mathbf{u}).
\end{align}
From Eqn.~\eref{eq:4b_Nsit}, the NS partial sum can be interpreted as an iterative update step, where the $i^{th}$ iteration refers to the $i^{th}$ partial sum.

\section{Optimization-based image reconstruction}
\label{subsec:4c}


A general approach to PACT image reconstruction is to formulate the sought-after estimate
of $p_0(\mathbf r)$ as the solution of an optimization problem.
  In fact, most modern image
reconstruction methods for computed imaging modalities including X-ray computed tomography
and magnetic resonance imaging are formulated in this way \cite{fessler1994penalized,wernick2004emission,pan2009commercial}.
There are potential practical and conceptual advantages to  optimization-based image reconstruction
over analytic methods.  
For example, because they are based on D-D forward models, optimization-based image reconstruction 
methods can comprehensively compensate for the imaging physics
and other physical factors such as responses of the measurement system that are not easily incorporated into an analytic
method.
Moreover, such methods provide a general framework for incorporating regularization, which
can mitigate the effects of measurement noise and  data incompleteness.
Because optimization-based methods are often implemented by use of iterative algorithms,
they are generally more computationally demanding than analytic methods; however the use of
modern parallel computing technologies \cite{wang2013accelerating} can render iterative three dimensional model-based reconstruction scheme for arbitrary photoacoustic acquisition geometries feasible for many
PACT applications.

Consider a D-D imaging model $\mathbf{u}=\mathbf{H} \boldsymbol{\theta}$ as described in Section~\ref{sec:section3}.
An optimization-based image reconstruction method seeks to determine an estimate of $\boldsymbol{\theta}$
 by solving an optimization program 
that can be generally specified as
\begin{align}\label{eq:4c_i_optim}
\hat{\boldsymbol{\theta}} = \underset{\boldsymbol{\theta}}{\textnormal{argmin}}{\ F(\mathbf{u},\boldsymbol{\theta}, \mathbf{H})}, \textnormal{ s.t.}\; f_i(\boldsymbol{\theta})  \in C_i,\ i = 1,\dots, N_c.
\end{align}
Here,  $F(\cdot)$ is the objective function to be minimized that depends
 on the D-D forward operator $\mathbf{H}$, the measured data $\mathbf{u}$,
 and the vector $\boldsymbol{\theta}$ that specifies the estimate of $p_0(\mathbf r)$.
 The functions $\{f_i(\boldsymbol{\theta})\}_{i = 0}^{N_c}$ and the closed set $\{C_i\}_{i = 0}^{N_c}$ describe  the set of $N_c$ constraints that the solution must satisfy.
It is important to note that numerous choices must be made in the design of the reconstruction method.
These include the specification of $\mathbf{H}$, $F(\cdot)$, and $f(\boldsymbol{\theta})$, as well as the
optimization algorithm that is employed to solve Eqn.\ (\ref{eq:4c_i_optim}).
It is the joint specification of these quantities that defines the sought after solution and
the numerical properties of the reconstruction method \cite{zhang2009effects}.
It should also be noted that many generic iterative  methods that
have been 
employed for PACT image reconstruction~\cite{paltauf2002iterative,AnastasioTATHT,haltmeier2017analysis}
can be interpreted as computing solutions to specific cases of
 Eqn.\ (\ref{eq:4c_i_optim}).
Below, some commonly employed optimization programs employed by image reconstruction methods are reviewed.

\subsection{Penalized least squares methods}
\label{subsubsec:PLS}




A special case of Eqn.\ (\ref{eq:4c_i_optim}) corresponds to a penalized least squared (PLS) estimator, and is given by
\begin{align}\label{eq:4c_ii_PLS}
\hat{\boldsymbol{\theta}} = \underset{\boldsymbol{\theta}\geq 0}{\textnormal{argmin}}\frac{1}{2} ||\mathbf{u} - \mathbf{H}\boldsymbol{\theta}||_{\mathbf{W}}^2 + \gamma R(\boldsymbol{\theta}). 
\end{align}
Since the initial pressure distribution is non-negative, the above problem
can be constrained so the solution satisfies $\boldsymbol{\theta}\geq 0$.
The objective function in the PLS estimator is expressed as two terms.  The quantity $\frac{1}{2}||\mathbf{u} - \mathbf{H}\boldsymbol{\theta}||_{\mathbf{W}}^2$ is referred to as the data fidelity term that corresponds to a 
weighted least squares functional.  
The matrix $\mathbf{W} \in \mathbb{R}^{QK \times QK}$  that defines the weighted $l_2$ norm is symmetric
and positive definite, whereas $\mathbf{u} \in \mathbb{R}^{QK \times 1}$ corresponds to the measurement vector.
This data fidelity functional
is convex  and differentiable with respect to $\boldsymbol{\theta}$.
For the case of Gaussian measurement noise, this data fidelity term can be interpreted
as a negative log-likelihood function; however, its use is not restricted to that case.

The quantity $R(\boldsymbol{\theta})$ is a penalty term that can be designed to regularize the inverse problem,
and $\gamma \in \mathbb{R}$ is a regularization parameter that controls the amount of regularization.
 A classic form of regularization is  Tikhonov regularization~\cite{golub1999tikhonov}, where the regularization term is specified as
\begin{align}
R(\theta) \equiv ||\boldsymbol{\theta}||_\mathbf{P}^2,
\end{align}
where $\mathbf{P} \in \mathbb{R}^{N \times N}$ is a symmetric positive definite matrix.
As the Tikhonov regularization term is differentiable with respect to $\boldsymbol{\theta}$, a variety
 of gradient-based methods can be employed to solve  Eqn.\ (\ref{eq:4c_ii_PLS}).

 Gradient-based methods that can be employed to solve Eqn.~\eref{eq:4c_ii_PLS} can be broadly grouped into two classes: first order methods and second order methods. As part of  first order methods, the Taylor approximation used to compute the descent direction involves a linear or first order approximation of the cost function. Hence, information about the gradient of the function is employed to compute the descent direction. Some of the most commonly employed first order methods include the steepest descent method, and the  conjugate gradient method. There also exists a class of first order methods that employ information about the past gradients/momentum to speed up the convergence rate of first order algorithms. Such methods are referred as Nesterov methods, whereby linear combinations of present and past gradients are used to compute the descent direction~\cite{nesterov1983method,nesterovapproach}. In addition, a family of methods called Krylov-based methods, the subset of which is are the conjugate gradient methods, are also employed to solve smooth, convex programs defined in Eqn.~\eref{eq:4c_ii_PLS}~\cite{paige1982lsqr,gutknecht2007brief,greenbaum1997iterative}. The design and study of variants of the Nesterov method and the Krylov-based methods to solve smooth, convex optimization programs is an active research area~\cite{ghai2019comparison,su2014differential,dax2019restarted}.

The second class of methods regularly employed to optimization programs are the second order methods. The Taylor approximation used to compute the descent direction for second order methods is a quadratic or second order approximation of the cost function. Hence, information about the inverse Hessian of the cost function is employed to compute the descent direction. Methods that explicitly compute the inverse Hessian are referred to as Newton methods. Although the newton methods have favorable converge properties, the computational burden associated with computing a Hessian and inverting it is prohibitively large. Hence, a variety of methods that approximate the inverse Hessian from first order gradient information are commonly employed. Such methods are referred to as Quasi-Newton methods. A variety of Quasi-Newton methods have been proposed and comprehensively studied~\cite{wright1999numerical,dennis1977quasi,nocedal1980updating,conn1991convergence,khalfan1993theoretical,dai2002convergence}.

The set of methods discusssed above can handle cost functions that have smooth regularization and data fidelity terms that are differentiable. However, certain modern approaches to regularization employ non-smooth choices for $R(\boldsymbol{\theta})$
based on the $l_1$-norm:
\begin{equation}
\label{eq:l1}
R(\theta)\equiv||\boldsymbol{\Phi}\boldsymbol{\theta}||_1,
\end{equation}
where  $\boldsymbol{\Phi} \in \mathbb{R}^{N \times N}$ is a sparsifying transformation that is chosen
such that $\boldsymbol{\theta}$ is sparse in its range.
Popular choices for $\boldsymbol{\Phi}$ include the discrete wavelet transform or
 finite difference operators. Total variation (TV) regularization \cite{sidky2008image} can be achieved as a special case
when the latter are employed.
Such choices are based on the observation that many objects possess a
 sparse representation in the wavelet or gradient domain \cite{starck2010sparse,chartrand2009fast}. 
Although Eqn.\ (\ref{eq:l1}) is convex, it is not differentiable with respect to $\boldsymbol{\theta}$. 
However, a variety of  non-smooth optimization  methods, such as proximal-gradient methods, can be employed to solve  Eqn.\ (\ref{eq:4c_ii_PLS}) in this case.
Proximal methods are a type of forward-backward splitting approach,
 which alternates between computing a gradient step on the data fidelity term
and a solution to a proximal problem involving the regularization
\cite{parikh2014proximal,combettes2011proximal,beck2009fast}. Moreover, the proximal gradient methods can also be accelerated by utilizing momentum information from past gradients or by deploying second order information to solve the proximal problem~\cite{becker2012quasi,beck2009fast,beck2009fasta,pan2013sparse,nesterov2007gradient,lee2012proximal,stella2017forward}.
\begin{figure}[h]
        \centering
        \subfigure[]{\includegraphics[width = 0.27 \linewidth,angle=90] {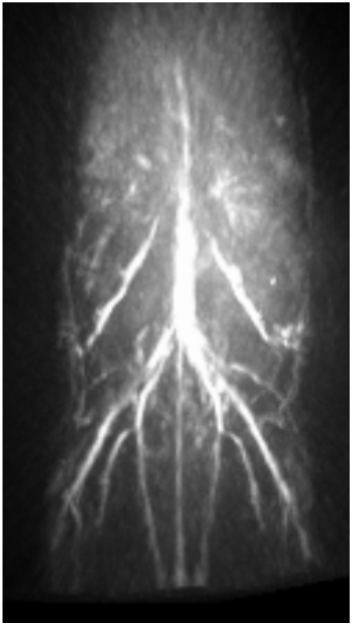}}
        \subfigure[]{\includegraphics[width = 0.27 \linewidth,angle=90] {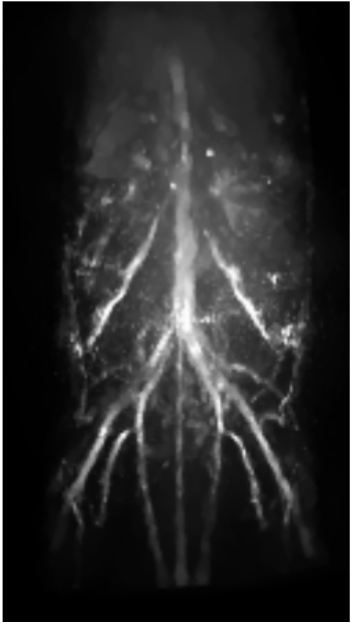}}
        \caption{Maximum intensity projection (MIP) renderings of 3D images of a live mouse  reconstructed by use of
the (a) FBP algorithm and (b) PLS estimator in Eqn.\ (\ref{eq:4c_ii_PLS})
that employed a TV penalty with $\gamma = 0.05$. Both the images are normalized to the same grayscale window. The images were reproduced from the  literature~\protect\cite{wang2012investigation}.}
        \label{fig:Mouse_3D}
\end{figure}

Examples of PACT images of a live mouse reconstructed
 by use of two different reconstruction methods are displayed in Fig.~\ref{fig:Mouse_3D}.
A description of the imaging system and experimental details have been described in previous works
\cite{ermilov2009development,brecht2009whole}. 
The small animal imager consisted of an arc-shaped probe of 64 transducers along the vertical axis. Thus, a tomographic view consisted of data recorded at 64 transducers along the arc. The complete tomographic dataset was acquired by rotating the object $360^{\circ}$ around the vertical axis. A 180-view dataset was used to reconstruct 3D images of the mouse trunk as shown in Fig.~\ref{fig:Mouse_3D}. 
 The image shown in Fig.~\ref{fig:Mouse_3D}(a) was reconstructed
by use of a FBP algorithm, while the image in Fig.~\ref{fig:Mouse_3D}(b)
was reconstructed by use of the PLS estimator in Eqn.\ (\ref{eq:4c_ii_PLS})
that employed a TV penalty.  
 The image corresponding to the PLS-TV estimate
possesses a higher contrast than the images reconstructed by use of
 the FBP algorithm and reveal a much sharper body vascular tree.
These observations are consistent with the fact  that optimization-based
image reconstruction methods can produce images that possess different
physical and statistical characteristics than the ones produced by analytical method.

\begin{figure}[htbp]
        \centering
        \subfigure[]{\includegraphics[width = 0.39 \linewidth] {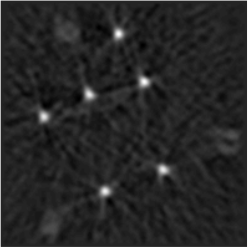}}
        \subfigure[]{\includegraphics[width = 0.40 \linewidth] {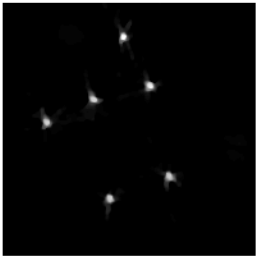}}
        \caption{2D slices of 3D images of an experimental phantom reconstructed from few view data (144 tomographic views) by the use of Eqn.\ (\ref{eq:4c_ii_PLS})  with
the penalty function specified as the (a) $l_2$-norm of a discrete  Laplacian and (b)  TV semi-norm.
  Both the images are of the same size and are normalized to the same grayscale window.The images were reproduced from the literature~\protect\cite{wang2012investigation}.}
        \label{fig:Tube_slice}
\end{figure}

An important consideration in the formulation of an optimization-based image
reconstruction method is the choice of the penalty function.
To demonstrate how different choices can influence image quality,
Fig.~\ref{fig:Tube_slice} displays images of a simple experimental phantom reconstructed by use of
 Eqn.\ (\ref{eq:4c_ii_PLS})  with two different  choices for the
penalty function. 
Figures~\ref{fig:Tube_slice} (a) and (b) correspond to use of
 quadratic smoothness Laplacian regularization and TV regularization, respectively.
 The quadratic Laplacian penalty corresponds to the $l_2$-norm
of a discrete Laplacian operator acting on the image estimate.
In this example, use of the TV penalty resulted in images with
reduced artifact levels and enhanced spatial resolution as
compared to use of the quadratic Laplacian penalty.

\begin{figure}[htbp]
        \centering
        \subfigure[]{\includegraphics[width = 0.45 \linewidth, height = 0.45 \linewidth] {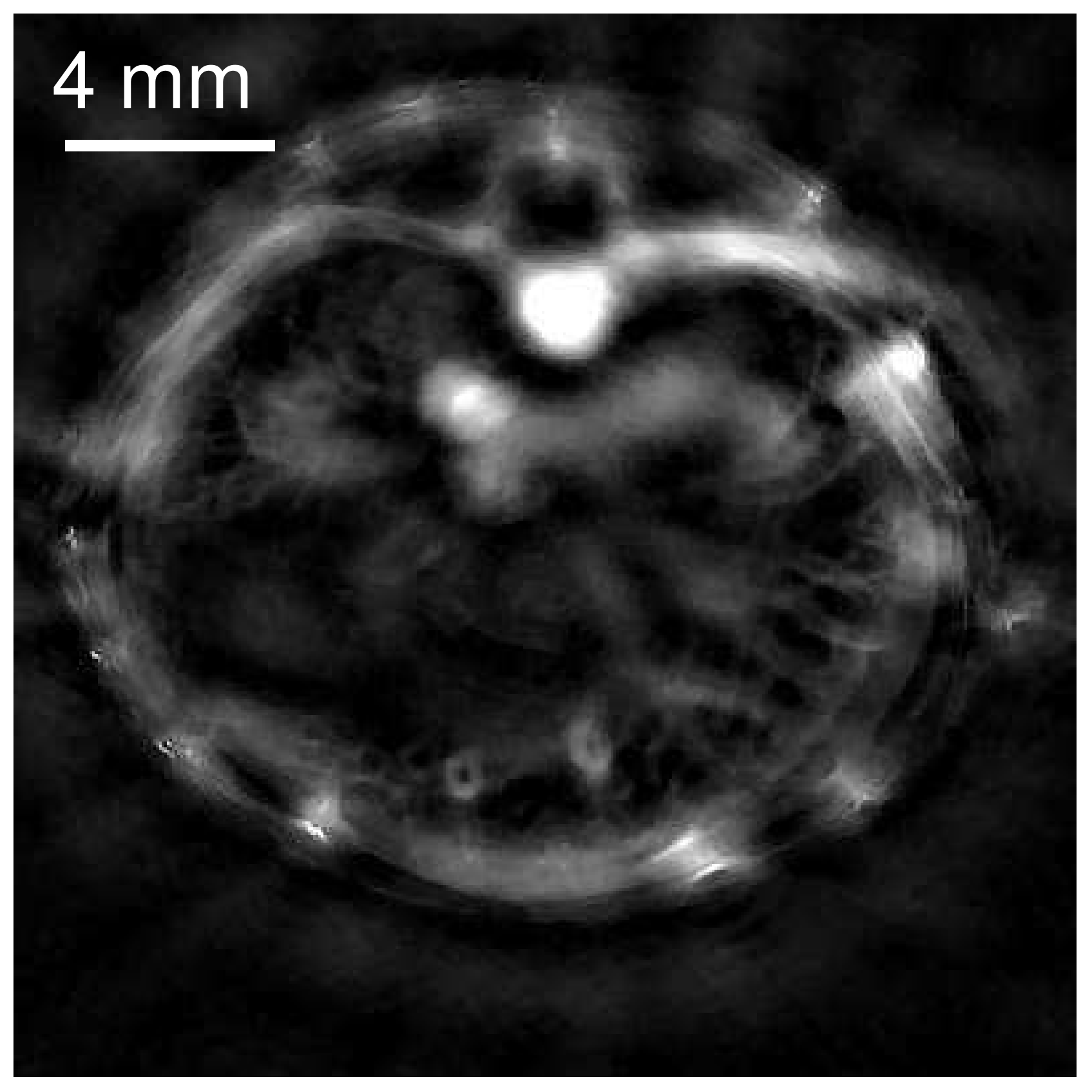}}
        \subfigure[]{\includegraphics[width = 0.45 \linewidth, height = 0.45 \linewidth] {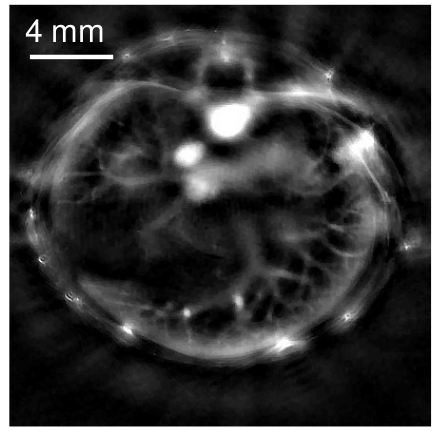}}
        \caption{2D reconstructed images of a mouse trunk produced by use of PLS-TV algorithm where (a) the imaging model compensated for the speed of sound variations between the background and the bulk tissue and (b) the imaging model used only one constant speed of sound value. The value of the regularization parameter,  $\gamma$ was set to be $0.01$. The images were reproduced from the literature~\protect\cite{matthews2018parameterized}.}
        \label{fig:mouse2d}
\end{figure}

All the images shown in Figures~\ref{fig:Mouse_3D} and~\ref{fig:Tube_slice} were reconstructed by use of an iterative image reconstruction model that assumed homogeneous acoustic media and thus could not account for the acoustic heterogeneity of the medium. However, knowledge about the spatial distribution of acoustic heterogeneity of the medium can be incorporated into an iterative reconstruction algorithm based on the imaging model described in Eqn.~\eref{eq:3c_HPS}. In vivo experimental studies have been conducted to validate the aforementioned iterative algorithm whereby the spatial distribution of the acoustic properties of the medium are accounted for in the reconstruction algorithm. For the in vivo studies, the experimental data were acquired from a mouse trunk using a small animal imaging system that has been described in detail in earlier works~\cite{li2017single}. As opposed to the previous studies where the medium was assumed to be homogeneous, in this study the acoustic variation between the background and the bulk mouse tissue were accounted for in the reconstruction algorithm.  A reconstructed initial pressure distribution image for fixed constant sound speed value is shown in Fig.~\ref{fig:mouse2d}(a).  In the reconstructed image, strong surface and interior vessel structures are observed. While the interior vessel structures appear out of focus, the surface vessels appear to be in focus. When assuming a homogeneous acoustic media, the image reconstruction algorithm could not produce images where both the surface and the interior vessels could be concurrently focused with a single tuned speed of sound value. To account for this, an image reconstruction algorithm that assigned different speed of sound values to the bulk tissue and the background was employed. The PLS-TV algorithm that was based on the imaging model defined in Eqn.~\eref{eq:3c_HPS} was employed to reconstruct the initial pressure distribution. The reconstructed initial pressure distribution when a heterogeneous speed of sound distribution is used is shown in Fig.~\ref{fig:mouse2d}(b). Comparing Figures~\ref{fig:mouse2d}(a) and~\ref{fig:mouse2d}(b), the surface and interior vessels are observed to be focused concurrently in the latter image, while only the surface vessels appear in focus in the former image. 

\subsection{Computation of adjoint operators and objective function gradients in PACT}

The reconstruction approaches described above are  general and applicable to a wide range of computational
inverse problems.  Here, PACT-specific details regarding the implementation of optimization-based
image reconstruction methods are reviewed.  Specifically, methods for computing the
adjoints of some of the D-D forward operators of Section~\ref{sec:section3} are presented. The explicit computation of the adjoint of the D-D forward operator facilitates  the computation of gradients of data fidelity functionals.

A necessary step in solving optimization problems of the form given in Eqn.\ (\ref{eq:4c_ii_PLS}) is the
computation of the gradient of the data fidelity  term
\begin{align}
\psi(\boldsymbol{\theta}) \equiv \frac{1}{2}||\mathbf{u} - \mathbf{H}\boldsymbol{\theta}||_{\mathbf{W}}^2.
\end{align}
with respect to $\boldsymbol{\theta}$.
Formally, this quantity is given by
\begin{align}
\nabla \psi(\boldsymbol{\theta}) = \mathbf{H}^{\dagger}\mathbf{W} (\mathbf{H}\boldsymbol{\theta} - \mathbf{u}),
\end{align}
where $\mathbf{H}^{\dagger}$ denotes the adjoint of the system matrix $\mathbf{H}$. As such, a key
step in computing the data fidelity gradient is computing the action of the adjoint operator.
As mentioned previously, for many problems of interest, $\mathbf{H}$ is too large to store in memory
and its action is commonly computed on-the-fly by use of an algorithm.
The same is true for $\mathbf{H}^{\dagger}$. Although the weighted $l_2$-norm is one of the most commonly employed data fidelity terms, a variety of convex data fidelity terms can be employed for PACT image reconstruction. Some, of the alternative convex data fidelity terms include weighted $l_1$-norm, KL-divergence, etc. Furthermore, the weight matrix associated with the weighted $l_2$-norm is a design parameter that can be constructed to incorporate \textit{a priori} information about the uncertainties associated with the imaging model. In cases where an arbitrary data fidelity term is employed, the gradient of the data fidelity term can be computed efficiently through use of the adjoint state method~\cite{norton1999iterative,plessix2006review}. In the adjoint state method, the gradient of the data fidelity term with respect to the model parameters are computed through the use of adjoint state variables. For PACT applications, the adjoint state variables are solutions to the adjoint of the wave equation defined in Eqn.~\eref{eq:2a_loss}. Hence, the computational complexity associated with the computation of the gradient of an arbitrary data fidelity term through the use of the adjoint state method would at least be on the same order as computing the action of the discrete adjoint operator $\mathbf{H}^{\dagger}$.

Although not discussed below, it should be noted that another option for establishing $\mathbf{H}^{\dagger}$ is to
employ an  \textit{adjoint-then-discretize approach}~\cite{arridge2016adjoint}.
 In such an approach, the explicit analytical form of the adjoint of the C-C forward operator is established.
Subsequently, the C-C adjoint operator is discretized to obtain an estimate $\mathbf{\hat H}^{\dagger}$ of
the D-D adjoint operator $\mathbf{H}^{\dagger}$.
From an implementation perspective, such approaches can sometimes be more convenient than 
computing the adjoint operator corresponding to the assumed D-D forward operator $\mathbf{H}$. However,
in cases where $\mathbf{\hat H}^{\dagger}$  does not accurately mimic $\mathbf{H}^{\dagger}$,
the behavior of iterative algorithms can be altered \cite{zeng2000unmatched}. The spectral properties of the forward-adjoint operator pair that govern convergence properties and the restrictions associated in choosing such operator pair have been studied~\cite{zeng2000unmatched,lou2018analysis}.

\subsubsection{Adjoint for interpolation-based forward model for homogeneous medium }
\label{subsec:4c_ii_3}

 Here, the interpolation-based D-D forward model described in Secs.\ \ref{subsec:3b_1} and \ref{subsubsec:3d_ii} is considered.
The matched adjoint operator corresponding to
 $\mathbf{H}_{int}$ in Eqn.~(\ref{eq:3d_int_DD}) is defined as $\mathbf{H}_{int}^\dagger = \mathbf{G}^\dagger \mathbf{D}^\dagger \mathbf{H}^{e\dagger}$, where
  $(\cdot)^\dagger$ denotes the transpose of the corresponding D-D operator (i.e., a matrix).
These operators can be  computed as \cite{wang2013accelerating}
\begin{gather}
    \mathbf{H}^{e\dagger} \mathbf{u} = \mathcal{F}^{-1}\{ \mathcal{F}(\mathbf{h}^e) \mathcal{F}(\mathbf{u}) \} \equiv \tilde{\mathbf{p}}_{ideal},           \label{eq:4c_ii_3_He_adjoint} \\
    [\mathbf{D}^\dagger \tilde{\mathbf{p}}_{ideal} ]_{qK+k} = \frac{\beta}{8 \pi C_p \Delta_t^2 k} ([\tilde{\mathbf{p}}_{ideal}]_{qK+k+1} - [\tilde{\mathbf{p}}_{ideal}]_{qK+k-1}) \equiv [\tilde{\mathbf{g}}]_{qK+k},      \label{eq:4c_ii_3_D_adjoint} \\
    [\mathbf{G}^\dagger \tilde{\mathbf{g}}]_{n} = \Delta_s^2 \sum_{q=0}^{Q-1} \sum_{k=0}^{K-1} [\tilde{\mathbf{g}}]_{qK+k} \sum_{i=0}^{N_i-1} \sum_{j=0}^{N_j-1} \phi_n(\mathbf{r}_{k, i, j}) \equiv [\tilde{\boldsymbol{\theta}}]_n.           \label{eq:4c_ii_3_G_adjoint}
\end{gather}

While the forward operator $\mathbf{H}_{int}$ can be efficiently
 implemented by use of parallel computing techniques \cite{wang2013accelerating},
 the adjoint operator $\mathbf{H}_{int}^{\dagger}$ is difficult to implement efficiently. 
This is partly due to the fact that when implementing $\mathbf{H}^\dagger$ in the interpolation-based model, it is difficult to satisfy the principle of ``partition on target results rather than sources" \cite{kirk2016programming} for safe and efficient GPU implementation. 
In addition, evaluating the backward operator $\mathbf{G}^{\dagger}$ relies on expensive atomic operations
 on the GPU \cite{wang2013accelerating}, resulting in a 6-10 times longer runtime for $\mathbf{H}_{int}^\dagger$ than for $\mathbf{H}_{int}$.

To address this problem, an approximation of the adjoint operator can be employed that closely
approximates the true adjoint but is computationally more efficient to compute~\cite{lou2018analysis}.
Such operators, denoted as $\mathbf{H}^\dagger_{unmatched}$,  are referred to as unmatched adjoint operators. 
An unmatched adjoint operator that approximates $\mathbf{H}_{int}^\dagger$
by use of a simplified voxel-driven model can be defined as: 
\begin{equation}
    \mathbf{H}_{unmatched}^\dagger \equiv \mathbf{G}_{unmatched}^\dagger \mathbf{D}^\dagger \mathbf{H}^{e\dagger},           \label{eq:4c_ii_3_H_unmatched} \\
\end{equation}
where
\begin{equation}
    [\mathbf{G}_{unmatched}^\dagger \tilde{\mathbf{g}}]_{n} = \Delta_s^2 \sum_{q=0}^{Q-1}[\tilde{\mathbf{g}}]_{\tilde{k}} \equiv [\tilde{\boldsymbol{\theta}}]_n.
\label{eq:4c_ii_3_G_unmatched}
\end{equation}
Here, $\tilde{k} \equiv (\frac{\mathbf{r}^s_q - \mathbf{r}_n}{c_0}) / \Delta_t$ and the value of $[\tilde{\mathbf{g}}]_{\tilde{k}}$ is approximated by linearly interpolating from its two neighboring samples: 
\begin{equation}
    [\tilde{\mathbf{g}}]_{\tilde{k}} \equiv (k+1-\tilde{k})[\tilde{\mathbf{g}}]_{qK+k} + (\tilde{k}-k)[\tilde{\mathbf{g}}]_{qK+k+1},  \nonumber
\end{equation}
with $k$ denoting the integer part of $\tilde{k}$. 
The operator $\mathbf{H}_{unmatched}^\dagger$ approximates the operation of $\mathbf{H}_{int}^\dagger$, but is much faster to compute. 
It can be seen from Eqns.~(\ref{eq:4c_ii_3_G_adjoint}) and (\ref{eq:4c_ii_3_G_unmatched}) that the computation of the exact adjoint $\mathbf{G}^\dagger$ involves $Q K N_i N_j \times \mathcal{O}(1)$ calculations, while the approximated adjoint $\mathbf{G}_{unmatched}^\dagger$ involves only $Q \times \mathcal{O}(1)$ calculations. 
Though it is difficult to express the computational complexity of the matched adjoint operator in strict big-\emph{O} notation, because $N_i$ and $N_j$ change with $q$ and $k$, the product of these two terms can be approximately on the order of 10,000 for 3D OAT studies;
 in such cases,  $Q K N_i N_j \times \mathcal{O}(1)$ is significantly larger than $Q \times \mathcal{O}(1)$. 

To demonstrate the use of the proposed unmatched adjoint operator, images were reconstructed
from experimental whole-body mouse PACT data by use of the PLS estimator in Eqn.\ (\ref{eq:4c_ii_PLS}) with a quadratic penalty term specified by the $l_2$ norm of the 3D gradient of the object. 
The 3D PACT dataset was acquired by use of a previously reported small animal imaging system~\cite{ermilov2009development}. 
   The system employed an arc-shaped transducer array containing
 64-elements that spanned 152 degrees over a circle of radius 65 mm. 
During scanning, the object was rotated over a full 360 degrees with 180 equispaced tomographic views. 
A Landweber iterative algorithm~\cite{landweber1951iteration} was employed to compute two different PLS estimates;
in one case, $\mathbf{H}_{int}^\dagger$ was employed and in the other $\mathbf{H}_{unmatched}^\dagger$ was employed.

\begin{figure}[htbp]
        \centering
    \includegraphics[width = 1.0 \linewidth]{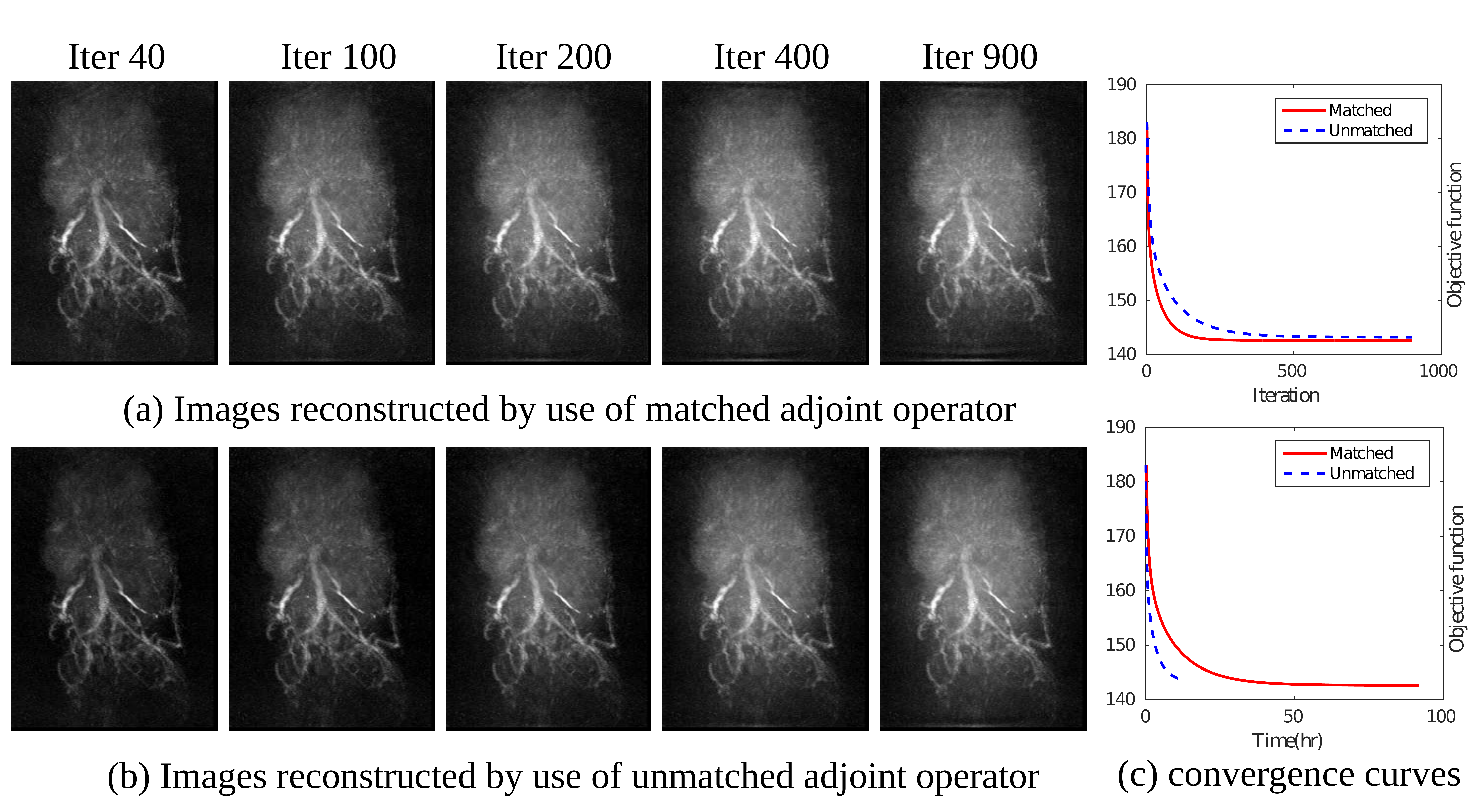}
    \caption
    {
        (a) and (b): Reconstructed 3D images of a mouse produced by use of matched and unmatched adjoint operators, with a regularization parameter of 0.001. All figures are MIP images.
        (c) The convergence curves of the Landweber iterative algorithm, where the top plot shows the objective function values against iterative number, while the bottom plot shows objective function value against computational time for the first 300 iterations. The images were reproduced from the literature~\protect\cite{lou2018analysis}.  }
   \label{fig:4c_ii_3_unmatched recon}
\end{figure}
The reconstructed PACT mouse images with regularization parameter $\lambda = 0.001$ are shown in Fig.~\ref{fig:4c_ii_3_unmatched recon}(a) and \ref{fig:4c_ii_3_unmatched recon}(b).
Each figure contains the maximum intensity projection (MIP) images of the reconstructed 3D volumes. 
The top and bottom rows contain the results corresponding to use
of  $\mathbf{H}_{int}^\dagger$ and $\mathbf{H}_{unmatched}^\dagger$, respectively.
The columns, from left to right, correspond to the reconstructed images after 40, 100, 200, 400, and 900 iterations, respectively. 
In addition, Fig.~\ref{fig:4c_ii_3_unmatched recon}(c) displays
 the objective function values versus iteration number (top row)
 and computational time (bottom row). 
These results demonstrate that, for this example,  use of $\mathbf{H}_{unmatched}^\dagger$ 
reduced computational times by approximately a factor of 6 while producing images
that are visually comparable in image quality to those obtained by use of $\mathbf{H}_{int}^\dagger$.

However, as mentioned above for the \textit{adjoint-then-discretize} approach, employing unmatched adjoint operators in iterative image reconstruction algorithms is not without risk.
Improperly designed unmatched adjoint operators may lead to algorithm divergence. 
In addition, the converged solution obtained by use of an unmatched adjoint operator will generally be different from the true solution. 
That being said, a carefully-designed unmatched operator can still be of great practical value for accelerating iterative reconstruction for large scale 3D problems~\cite{lou2018analysis}.


\subsubsection{Adjoint for full-wave forward model}
\label{subsec:4c_ii_4}

Here, the adjoint operator corresponding to the full-wave  D-D forward model  described in Section~\ref{sec:DDPS} for use with known heterogeneous acoustic media is considered.
 Given the definition of $\mathbf{H}_{PS}$ in Eqn.~\eref{eq:3c_HPS}, the adjoint of $\mathbf{H}_{PS}$ can be computed as 
\begin{align}\label{eq:4c_ii_4_adj_PS}
\mathbf{H}_{PS}^{\dagger} = \mathbf{T}_{0}^{\dagger} \mathbf{T}_1^{\dagger} \cdots \mathbf{T}_{K - 1}^{\dagger} \mathbf{M}^{\dagger},
\end{align}
where the adjoint of the temporal propagator matrices $\mathbf{T}_k \in \mathbb{R}^{7NK \times 7NK}$
 can be derived from Eqn.~\eref{eq:2a_loss}~\cite{huang2013full}. The adjoint of the forward operator described in Eqn.~\eref{eq:4c_ii_4_adj_PS} can be interpreted as an explicit reversal of the computational steps of the forward scheme~\cite{huang2013full}. \\

\section{Joint reconstruction of initial pressure distribution and acoustic medium parameters}
\label{sec:section5}

The forward models and image reconstruction methods surveyed above
assume either that (1) the to-be-imaged object and surrounding medium are
acoustically homogeneous or (2) the object and coupling medium are
acoustically heterogeneous  and the spatially variant acoustic parameters 
are known.
However, it remains generally
 difficult and/or inconvenient to accurately estimate the spatially variant
acoustic parameters.  Accordingly, the vast majority of reported PACT
studies assume that the medium is acoustically homogeneous and tune the values of the spatially invariant acoustic parameters to mitigate the impact of acoustic aberrations
on the reconstructed image.  

Below, a non-conventional approach to PACT image reconstruction is reviewed
in which $p_0(\mathbf r)$ and the spatially variant distribution of the
acoustic parameters are jointly estimated. This is referred
to as a joint reconstruction (JR) approach for PACT \cite{zhang2006reconstruction,jiang2006spatially}.
A physical motivation for attempting JR results from the observation that
  spatial variations in the acoustic parameters induce aberrations
 in the photoacoustic wavefields; consequently, the measured PACT data encodes
 some information about the acoustic parameters.
 Several JR methods have been proposed in which the initial pressure distribution and
 speed of sound (SOS) distribution are estimated concurrently from PACT data alone~\cite{zhang2008simultaneous,yuan2006simultaneous,chen2013tr,kirsch2012simultaneous,huang2016joint}. 

Motivated by the earlier numerical investigations, mathematical studies of the JR problem
 have been
 conducted~\cite{liu2015determining,hickmann2010unique,kirsch2012simultaneous,stefanov2012instability}. These studies established that,  when neglecting the discrete sampling effects, the initial pressure distribution and the SOS distribution can be uniquely determined from the measured PACT data only under certain restrictive
 assumptions \cite{liu2015determining}.
 In addition,  for a linearized version of the JR problem, it was
demonstrated that the SOS distribution and $p_0(\mathbf r)$
 could not be stably recovered from PACT data alone~\cite{stefanov2012instability}. For a linearized version of the forward problem, where the linearization is represents a smoothing operator for $p_0(\mathbf{r})$ and $c(\mathbf{r})^2$, the JR problem is unstable in any scale of Sobolev spaces~\cite{stefanov2012instability}. This suggests that solving the JR problem from PACT data alone where the imaging model is described by Eqn.~\ref{eq:2a_HCC} is an extremely challenging undertaking. Various approaches have been proposed to overcome these challenges. A possible approach to mitigate the problem is to augment the PACT measurements with a sparse set of ultrasound tomography data~\cite{matthews2017joint}.

Below, based on the D-D forward model for use with heterogeneous
acoustic media, a general
formulation of the JR problem in a discrete setting is provided.
Subsequently, a 
low-dimensional  parameterized version of the JR problem is reviewed that holds
promise for practical applications.

\subsection{JR algorithm}

Here, a JR problem is considered in which the spatially variant SOS distribution
of the object and coupling medium are unknown, but all other acoustic parameters
are known.  Let $\mathbf{c} \in \mathbb{R}^{N \times 1}$ denote a
finite-dimensional representation of the SOS distribution, the explicit
form of which will depend on the choice of numerical scheme that is adopted to
solve the wave equation in Eqn.~\eref{eq:2a_loss}.

Consider the D-D PACT forward model in Eqn.\ (\ref{D_D:dis_to_dis_img_mod})
that is re-stated as
\begin{align}
	\mathbf{u} = \mathbf{H}(\mathbf{c}) \boldsymbol{\theta},
\end{align}
where $\mathbf{H}(\mathbf{c})\in \mathbb{R}^{QK \times N}$ denotes a D-D forward model 
for use with heterogeneous acoustic media as described in Section~\ref{sec:DDPS}.
The notation $\mathbf{H}(\mathbf{c})$
 is employed to  make explicit
the dependence of forward operator on the unknown discretized
SOS distribution $\mathbf{c}$.

 The JR problem can be formulated as 
\begin{align}\label{eq:5_JR}
	\hat{\boldsymbol{\theta}}, \hat{\mathbf{c}} = \underset{\boldsymbol{\theta} \geq 0, \mathbf{c} > 0}{\textnormal{argmin}}\ \frac{1}{2} ||\mathbf{u} - \mathbf{H}(\mathbf{c})\boldsymbol{\theta}||_{\mathbf{W}}^2 + \beta_1 R_{c}(\mathbf{c}) + \beta_2 R_{\theta}(\boldsymbol{\theta}),
\end{align}
where $R_{c}(\mathbf{c})$ and $R_{\theta}(\boldsymbol{\theta})$ are convex penalty
 functions whose relative weights are determined
by the regularization parameters  $\beta_1$ and $\beta_2$, respectively.

To solve the JR problem in Eqn.~\eref{eq:5_JR},
 an alternating minimization optimization can be employed as described in  
Algorithm~\ref{alg:5_ALT} in which two subproblems are solved alternatively until a convergence condition is 
satisfied~\cite{huang2016joint} . The two subproblems can be expressed as 
\begin{align}
	\textnormal{Subproblem 1:  }&\hat{\boldsymbol{\theta}} = \underset{\boldsymbol{\theta} \geq 0}{\textnormal{argmin}}\ \frac{1}{2} ||\mathbf{u} - \mathbf{H}(\mathbf{c})\boldsymbol{\theta}||_{\mathbf{W}}^2 + + \beta_2 R_{\theta}(\boldsymbol{\theta})\\
	\textnormal{Subproblem 2: }& \hat{\mathbf{c}} = \underset{ \mathbf{c} > 0}{\textnormal{argmin}}\ \frac{1}{2} ||\mathbf{u} - \mathbf{H}(\mathbf{c})\boldsymbol{\theta}||_{\mathbf{W}}^2 + \beta_1 R_{c}(\mathbf{c}).\label{eq:5_SP2}
\end{align}
For a fixed  $\mathbf{c}$, Subproblem 1 corresponds to the conventional PACT reconstruction
problem stated in Eqn.\ (\ref{eq:4c_ii_PLS}), which is a convex optimization problem since the objective function
is convex for fixed $\mathbf{c}$.
 In Subproblem 2, $\boldsymbol{\theta}$ is fixed and an estimate of
the SOS distribution $\mathbf{c}$ is determined.  This corresponds to a non-convex problem.
When solving Eqn.~\eref{eq:5_SP2} by use of gradient-based methods,  the computation of the gradient of the data fidelity term
 with respect to the model parameter $\mathbf{c}$
 can be accomplished by use of the adjoint state
 method~\cite{norton1999iterative,plessix2006review}.

\begin{algorithm} \caption{Alternating optimization approach to JR of $\boldsymbol{\theta}$ and $\mathbf{c}$}
	\begin{algorithmic}[1]
		\renewcommand{\algorithmicrequire}{\textbf{Input:}}
		\renewcommand{\algorithmicensure}{\textbf{Output:}}
		\Require $\boldsymbol{\theta}^0, \mathbf{c}^0, \epsilon_{\theta}, \epsilon_{c}, \beta_1, \beta_2$.
		\Ensure  $ \hat{\boldsymbol{\theta}}, \hat{\mathbf{c}}$ 
		\State k $=$ 0, {k is the iteration number}. 
		\While{$\epsilon_{\theta} < {\epsilon^{\theta}_F}$ and $\epsilon_{c} < \epsilon^{c}_F$} \do \\
		\State $i_k = $ 0.
		\State $\boldsymbol{\theta}^{k + 1} = \underset{\boldsymbol{\theta^k} \geq 0}{\textnormal{argmin}}\ \frac{1}{2} ||\mathbf{u} - \mathbf{H}(\mathbf{c}^k)\boldsymbol{\theta^k}||_{\mathbf{W}}^2 + + \beta_2 R_{\theta}(\boldsymbol{\theta}^k)$.
		\State $\mathbf{c}^{k + 1} = \underset{ \mathbf{c}^k > 0}{\textnormal{argmin}}\ \frac{1}{2} ||\mathbf{u} - \mathbf{H}(\mathbf{c}^k)\boldsymbol{\theta}^{k+1}||_{\mathbf{W}}^2 + \beta_1 R_{c}(\mathbf{c}^k)$. 
		\State $\epsilon^{\theta}_F = \frac{|\boldsymbol{\theta}^{k+1} - \boldsymbol{\theta}^k|}{\boldsymbol{\theta}^{k+1}}$
		\State $\epsilon^{c}_F = \frac{|\boldsymbol{c}^{k+1} - \boldsymbol{c}^k|}{\boldsymbol{c}^{k+1}}$
		\State k $=$ k + 1, 
		\EndWhile 
		\State $ \hat{\boldsymbol{\theta}}\leftarrow \boldsymbol{\theta}^{(k)}$, $\hat{\mathbf{c}} \leftarrow \mathbf{c}^k$
	\end{algorithmic} 
	\label{alg:5_ALT}
\end{algorithm}

Since Subproblem 2 is non-convex, obtaining its accurate solution represents a challenge. In the subsequent section, some insights obtained from computer-simulation studies that shed a light on the numerical instability and the non-uniqueness properties of the JR problem will be reviewed.
\subsubsection{Numerical instability of JR methods}
\begin{figure}[htp]
	\centering
	\subfigure[]{\includegraphics[width = 0.40 \linewidth] {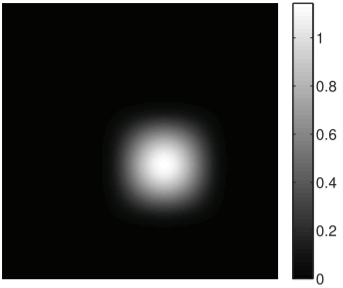}}
	\subfigure[]{\includegraphics[width = 0.40 \linewidth] {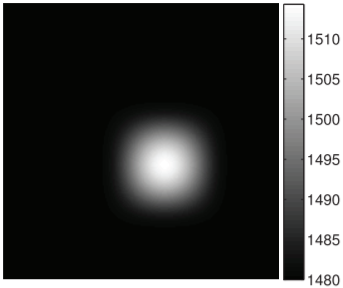}}
	\subfigure[]{\includegraphics[width = 0.40 \linewidth] {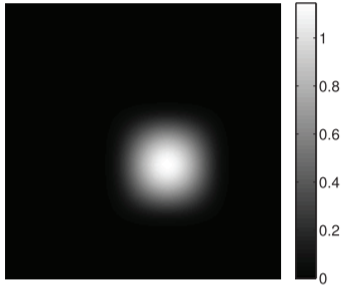}}
	\subfigure[]{\includegraphics[width = 0.40 \linewidth] {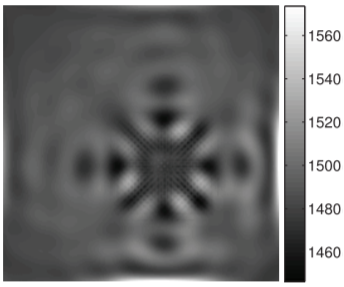}}
	\caption{Effect of perturbation of $\boldsymbol{\theta}$. Two numerical phantoms representing $\boldsymbol{\theta}$ are shown in subfigures (a) and (c). The two phantoms are very similar with a RMSE between them of only 0.004. Unregularized estimates of $\mathbf{c}$ reconstructed  by use of noiseless simulated measurement data and $\boldsymbol{\theta}$ specified in (a) and (c) are shown in subfigures (b) and (d), respectively. The images were reproduced from the literature~\protect\cite{huang2016joint}.}\label{fig:5_JR_instability}
\end{figure}

\begin{figure}[htp]
	\centering
	{\includegraphics[width = 1.20 \linewidth] {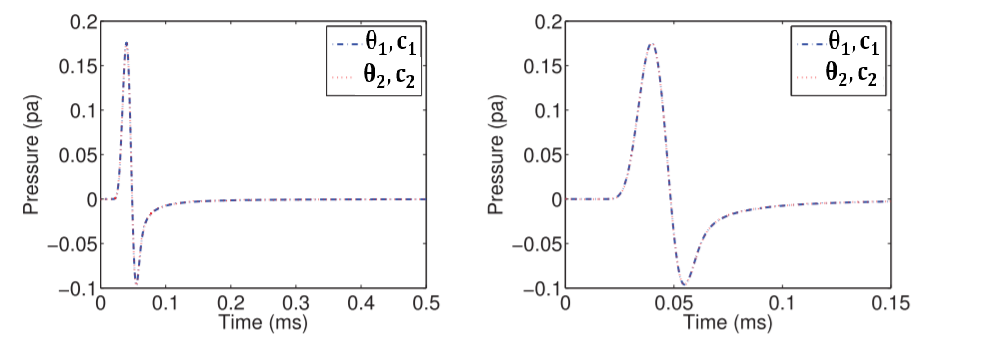}}
\caption{Numerical evidence of non-uniqueness of the JR problem: Simulated PA measurement data were computed from the $(\boldsymbol{\theta},\mathbf{c})$ pairs shown in Figures~\ref{fig:5_JR_instability}(a) and~\ref{fig:5_JR_instability}(b) and Figures~\ref{fig:5_JR_instability}(c) and~\ref{fig:5_JR_instability}(d). The two pressure profiles corresponding to an arbitrary transducer location are superimposed in both figures. The figure on the right, displays a zoomed-in version of the figure on the left. Similar agreement between the profiles was observed at all transducer locations. The images were reproduced from the literature~\protect\cite{huang2016joint}.}\label{fig:5_JR_nonuniqueness}
\end{figure}

To explore the numerical instability associated with solving Subproblem 2, computer-simulation studies were performed to provide insights into how small perturbations in the assumed $\boldsymbol{\theta}$ affects the accuracy of the reconstructed $\mathbf{c}$~\cite{huang2016joint}. Figures~\ref{fig:5_JR_instability}(a) and~\ref{fig:5_JR_instability}(c) display two similar normalized numerical phantoms
depicting $\boldsymbol{\theta}$. The RMSE between these phantoms is 0.004.
Simulated ideal PACT measurements were computed corresponding
to the cases where each of the two  $\boldsymbol{\theta}$ was paired
with a distinct SOS distribution $\mathbf{c}$. The utilized $\mathbf{c}$ are not shown
but is nearly identical to the reconstructed image shown in
Fig.~\ref{fig:5_JR_instability}(b). It is important to note that the $(\boldsymbol{\theta},\mathbf{c})$ pair in
the top row of Fig.~\ref{fig:5_JR_instability} produces nearly identical PA data to that produced by the $(\boldsymbol{\theta},\mathbf{c})$ pair in the
bottom row. The simulated noiseless normalized pressure data at an arbitrary
transducer location produced by the two $(\boldsymbol{\theta},\mathbf{c})$ pairs are
shown in Fig.~\ref{fig:5_JR_nonuniqueness}. The normalized pressure signals in Fig.~\ref{fig:5_JR_nonuniqueness} are observed to
overlap almost completely and the RMSE between the two sets of
PA data was $3.2e-4$.
Figures~\ref{fig:5_JR_instability}(b) and~\ref{fig:5_JR_instability}(d)  display the reconstructed estimates of
$\mathbf{c}$ when the $\boldsymbol{\theta}$ specified in Fig.~\ref{fig:5_JR_instability}(a)  and~\ref{fig:5_JR_instability}(c) was assumed,
respectively. These results demonstrate that the problem of
reconstructing $\mathbf{c}$ for a given $\boldsymbol{\theta}$ is ill-posed in the sense
that small changes in $\boldsymbol{\theta}$ can produce significant changes in the
reconstructed estimate of $\mathbf{c}$. This observation is consistent with
the theoretical results~\cite{stefanov2012instability}.


\begin{figure}[htp]
	\centering
	\subfigure[]{\includegraphics[width = 0.40 \linewidth] {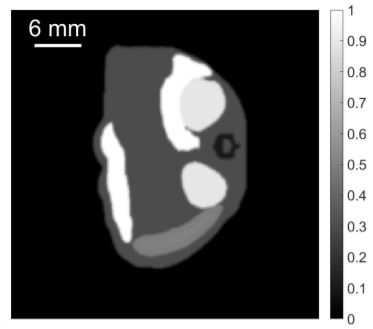}}
	\subfigure[]{\includegraphics[width = 0.40 \linewidth] {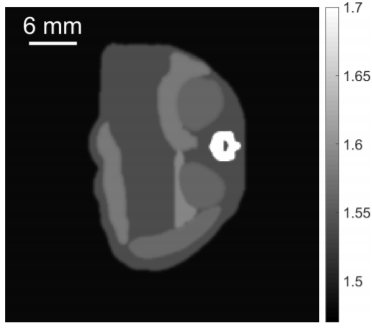}}
	\caption{Phantoms for (a) normalized initial pressure distribution $\boldsymbol{\theta}$, given in arbitrary units, (b) the sound speed distribution $\mathbf{c}$, given in units of $\frac{mm}{\mu s}$. The images were reproduced from the literature~\protect\cite{matthews2018parameterized}.} \label{fig:5_Phantom}
\end{figure}

\begin{figure}[htp]
	\centering
	\subfigure[]{\includegraphics[width = 0.40 \linewidth] {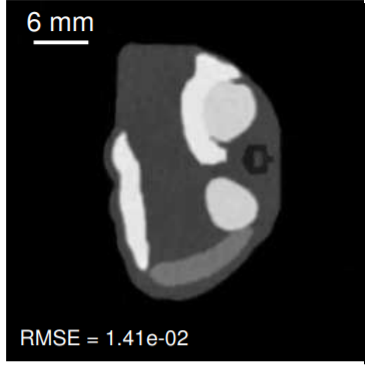}}
	\subfigure[]{\includegraphics[width = 0.40 \linewidth] {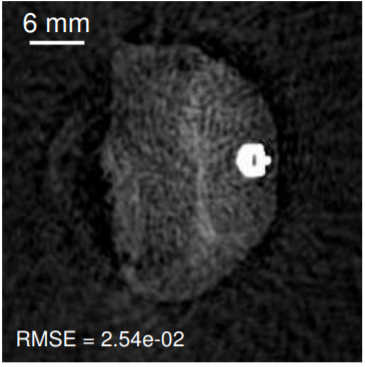}}
	\caption{ Jointly reconstructed (a) initial pressure distribution $\boldsymbol{\theta}$, given in arbitrary units, and (b) the sound speed distribution $\mathbf{c}$, given in units of $\frac{mm}{\mu s}$. The grayscale windows employed for displaying Figures~(a) and (b) correspond to the grayscale windows used in Figures~\ref{fig:5_Phantom}(a) and~\ref{fig:5_Phantom}(b), respectively. The images were reproduced from the literature~\protect\cite{matthews2018parameterized}.} \label{fig:5_Phantom_recon}
\end{figure}

Due to the non-convex nature of Subproblem 2, the optimization algorithm can return a local minimum (i.e., an inaccurate JR solution) as opposed to a global minimum (i.e., an accurate JR solution). Even though the estimate of $\mathbf{c}$ obtained at the local minimum is inaccurate, an accurate estimate of the initial pressure distribution $\boldsymbol{\theta}$ can sometimes be obtained. Various numerical studies have been conducted to show the non-uniqueness properties of the JR algorithm. Figures~\ref{fig:5_Phantom}(a) and~\ref{fig:5_Phantom}(b) display two numerical phantoms that describe the initial pressure distribution and the speed of sound distribution of a mouse phantom, respectively. The JR algorithm was applied to the forward PACT measurement data generated from the corresponding phantoms. The alternating minimization algorithm described in Algorithm~\ref{alg:5_ALT} was used to solve the JR algorithm. The results from the JR algorithm are displayed in Fig.~\ref{fig:5_Phantom_recon}. From the reconstruction results, one can observe that even though the sound speed distribution is highly inaccurate, the initial pressure distribution can be effectively estimated. This demonstrates that JR algorithms can be employed to improve PACT image quality even if the produced estimates of the sound speed distribution is highly inaccurate.
\subsection{Parameterized JR}
To mitigate the instability of JR,
a modified version of the JR problem can be formulated
in which $p_0(\mathbf r)$ and only  a lower  dimensional estimate of  SOS distribution  
are estimated \cite{zhang2008simultaneous,matthews2018parameterized,poudel2018joint}.
The SOS representation $\mathbf{c}$ is typically formed by use of pixel
expansion functions. To constrain the JR problem,
 a lower dimensional parameterized representation of
 the SOS distribution, $\mathbf{c}_p \in \mathbb{R}^{D}$, can be introduced as 
\begin{align}\label{eq:5_Parameter}
	\mathbf{c} =  \boldsymbol{\Gamma} \mathbf{c}_p,
\end{align}  
where $\boldsymbol{\Gamma} \in \mathbb{R}^{N \times D}$ is a binary matrix that maps
 $\mathbf{c}_p$ to  $\mathbf{c}$.
Let $\mathcal{I}_j$ denote the set of pixel indices in $\mathbf{c}$
 that corresponds to the $j^{th}$-parameterized SOS value (i.e., component
of $\mathbf{c}_p$),
  $\boldsymbol{\Gamma}$ can be defined as
\begin{align}\label{eq:5_gamma}
	[\boldsymbol{\Gamma}]_{i,j} = \begin{cases}
		1,\ \ \ &i \in \mathcal{I}_j \\
		0,\ \ \ &\textnormal{otherwise}
		\end{cases}.
\end{align}

 The parameterized JR problem is given by \cite{matthews2018parameterized}
\begin{align}
\label{eq:parameterizedJR}
	\hat{\boldsymbol{\theta}}, \hat{\mathbf{c}_p} = \underset{\boldsymbol{\theta} \geq 0, \mathbf{c}_p > 0}{\textnormal{argmin}}\ \frac{1}{2} ||\mathbf{u} - \mathbf{H}(\boldsymbol{\Gamma}\mathbf{c}_p)\boldsymbol{\theta}||_{\mathbf{W}}^2 + \beta_2 R_{\theta}(\boldsymbol{\theta}).
\end{align}
Once $\mathbf{c}_p$ is estimated, an estimate of $\mathbf{c}$
can be obtained by use of Eqn.\ (\ref{eq:5_Parameter}).
The gradient of the cost function with respect to $\mathbf{c}_p$ can be related to the gradient with respect to $\mathbf{c}$ via the chain rule.
 When $\Gamma$ corresponds to a linear mapping, one obtains
\begin{align}
	\nabla_{\mathbf{c}_p}\Big( \frac{1}{2} ||\mathbf{u} - \mathbf{H}(\boldsymbol{\Gamma}\mathbf{c}_p)\boldsymbol{\theta}||_{\mathbf{W}}^2 \Big) = \Gamma^{\dagger}\nabla_{\mathbf{c}}\Big(\frac{1}{2} ||\mathbf{u} - \mathbf{H}(\mathbf{c})\boldsymbol{\theta}||_{\mathbf{W}}^2\Big),
\end{align}
\begin{figure}[htp]
        \centering
        \subfigure[]{\includegraphics[width = 0.40 \linewidth] {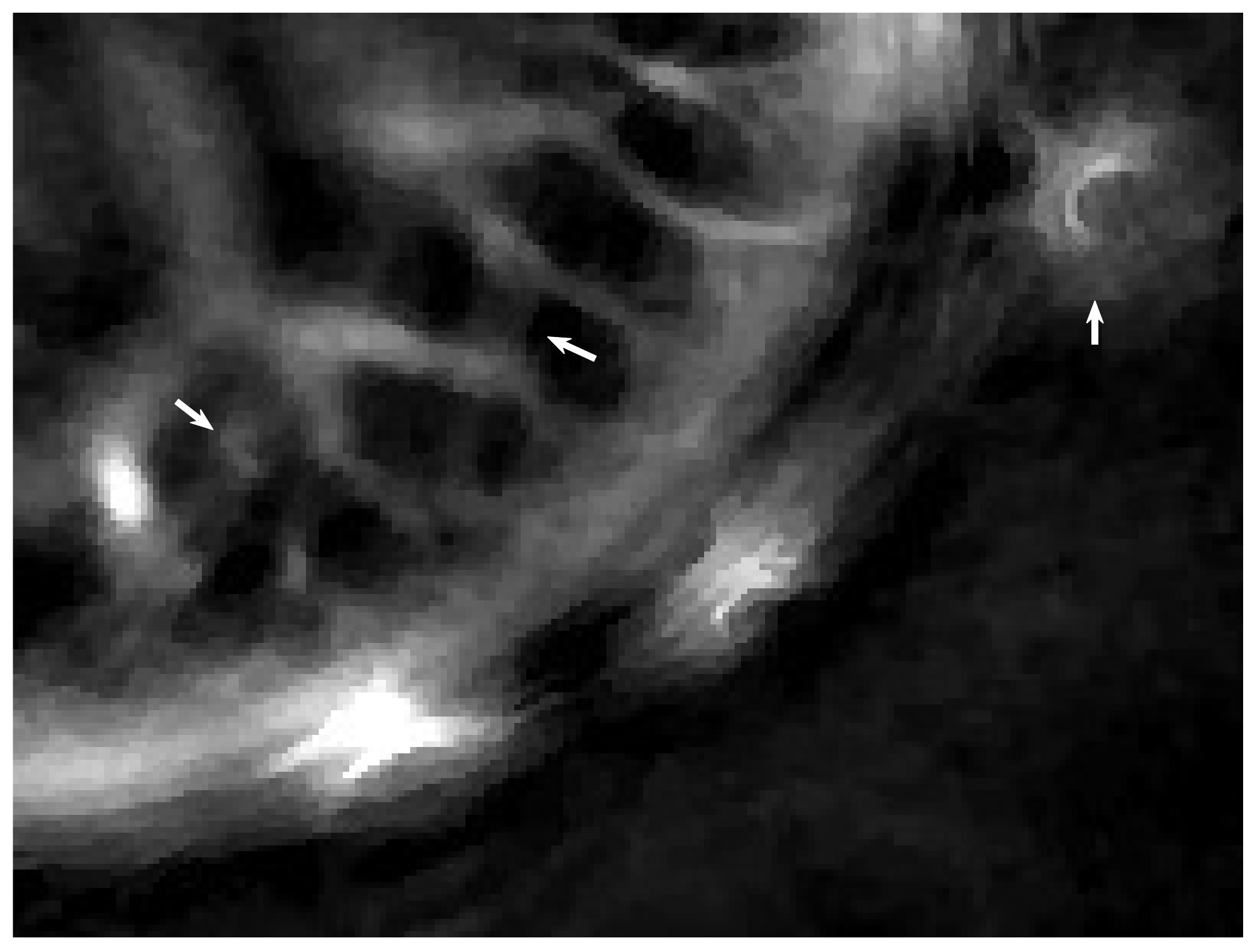}}
        \subfigure[]{\includegraphics[width = 0.40 \linewidth] {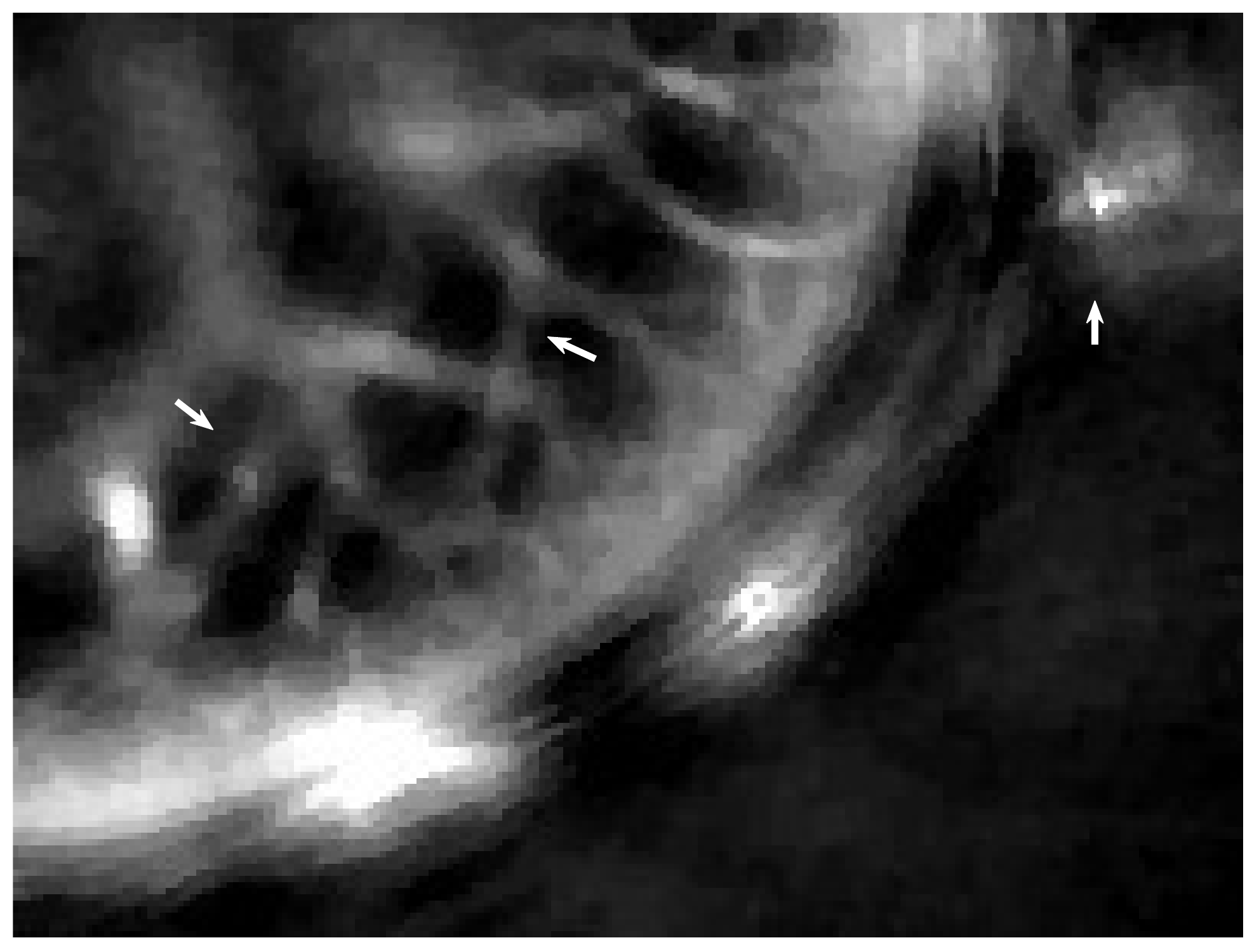}}
        \caption{Zoomed-in region of the reconstructed $p_0(\mathbf r)$ assuming  (a) a tuned constant sound speed of 1.500~mm/$\mu$s and (b) the SOS distribution obtained by parameterized JR. The arrows point to structures that are in focus in the JR image, but not in the tuned constant sound speed image. Results are shown in a grayscale window of $[0, 6000]$. The images were reproduced from the literature~\protect\cite{matthews2018parameterized}.} \label{fig:mouse_reconjr_zoom}
\end{figure}
where the gradient of the cost function with respect to $\mathbf{c}$ can
be computed by use of the adjoint state method~\cite{plessix2006review}.
Subsequently, an algorithm similar to Algorithm~\ref{alg:5_ALT} can be employed for image
reconstruction, where $\mathbf{c}$ is replaced by $\mathbf{c}_p$.
Alternatively, a weighted proximal gradient method has been proposed
to solve Eqn.\ (\ref{eq:parameterizedJR}) in which the necessary gradients can be computed
with only two wave solver runs \cite{matthews2018parameterized}.

To demonstrate the use of the parameterized JR algorithm, images were reconstructed from experimental mouse data using the PACT imager described earlier~\cite{li2017single}. Figure \ref{fig:mouse_reconjr_zoom} displays zoomed-in regions of images of
a live mouse reconstructed by use of an iterative method that employed
a constant SOS value of  1.500~mm/$\mu$s   (subfigure (a))
 and a JR method that employed a two-parameter SOS model
(subfigure (b)) \cite{matthews2018parameterized}.  To establish the two-parameter SOS model, the outer boundary of the mouse was manually segmented from the estimate of $p_0(\mathbf r)$ that was reconstructed
using the fixed SOS value.
The simulation grids consisted of $2048\times2048$ pixels with a pixel size of 0.05 mm. 
In the JR method, the initial guess for the parameterized sound speed distribution was 1.48~mm/$\mu$s for the background and 1.54~mm/$\mu$s for the mouse, while the initial
guess for $\boldsymbol{\theta}$ was the zero-vector. 
The JR result  demonstrates improvement
 over the image obtained with the constant SOS.
 The largest difference can be seen in the rightmost surface vessel, which appears as an arc in the constant sound speed image and a point in the JR image. In addition, several interior vessels are better focused in the JR image.

\section{Other challenges}
\label{sec:section6}
There are numerous other challenges and opportunities for
research related to image reconstruction for practical
applications of PACT that
are beyond the scope of this article.  Some of these are
outlined below.
\subsection{Approximating PACT as a 2D problem}
\label{sec:7a}

In practice, PACT imaging systems are often designed to
image 2D sections through a 3D object. This can be accomplished
by employing an array of elevationally focused ultrasound transducers.
For example, several groups have developed systems that employ
a circular ring array (or part of a ring) that surrounds the
2D section to be imaged \cite{xia2014small,li2017single,gamelin2009real,ma2009multispectral,lin2018single}. 
In such systems, a 3D object can be imaged on a 2D slice-by-slice basis
and the resulting images stacked together along the direction perpendicular to the
plane of the array to estimate a 3D reconstructed volume.
There are practical advantages to such approaches.
From a hardware perspective, 2D systems can be less costly to construct
than full 3D systems and may not require mechanical scanning of the transducer
array to image a 2D slice.   When advanced electronics are employed,
2D systems can yield near real-time data-acquisition; this can result in
excellent  temporal resolution and spatial resolution that is not
strongly compromised by physiological motion.
Image reconstruction can also be performed in near real-time \cite{buehler2010video}.


However, from an image reconstruction perspective,  there are potential 
challenges associated with such 2D approaches to PACT associated with the
fact that PACT is inherently a 3D technique.
If (idealized) focused transducers could be employed in a PACT imager
 to isolate an infinitesimally thin slice of the object,
 the spherical Radon transform forward model described in Section \ref{subsec:2b}
 would reduce to
 a circular Radon transform \cite{wang2008elucidation}. In this case, the
 reconstruction problem could be treated in 2D.
 However, the focused piezoelectric transducers employed in practice
do not perfectly reject acoustic signals that arrive from outside the plane of interest.
Moreover, the focusing properties of such transducers possess a temporal frequency dependence as described in Section \ref{subsubsec:3d_ii}.
Such factors and the fact that the PA wavefield propagates
according to a 3D wave equation
 result in a mismatch between the measured data and a canonical  2D PACT forward model.
Finally, when 2D PACT images corresponding to different parallel
slices of the object
are stacked together to estimate a 3D reconstructed volume, the
resolution is generally severely compromised 
in the direction perpendicular to the
plane of the array. The presence of artifacts in 2D PACT images due to the presence of 3D out-of-plane scatterers can also significantly degrade image quality. Similar issues have been studied in the geophysics community and medical ultrasound computed tomography communities~\cite{auer2013critical,yedlin2012uniform,bleistein2012mathematical,wiskin20173,duric2014clinical,gemmeke20173d,huthwaite2011high}. Experienced practitioners can help mitigate some of these out of plane scattering artifacts by carefully designing the elevational focusing properties
of the transducers~\cite{duric2014clinical,lin2018clinical,li2017single}.
	
The discrete forward models and optimization-based image reconstruction methods 
reviewed in the previous sections can address these issues and potentially improve the accuracy
and spatial resolution of images reconstructed by use of ring-array systems.
Namely, to compensate for the transducer focusing effects,
the SIRs of the transducers can be incorporated into a 3D discrete
forward model. In practice, however, it may be challenging to accurately
model the SIR for the case where an acoustic lens is employed. By use of a 3D forward model,  $p_0(\mathbf r)$ within a thin 3D volume can be estimated.  
In such approaches, parameters such as voxel size must be designed carefully to balance the effects of
data-incompleteness and model error.

\subsection{Spatiotemporal image reconstruction}

The majority of PACT reconstruction algorithms assume static imaging conditions;
namely, the object of interest is considered to be fixed and independent of time.
 This assumption, however, is violated in many  biomedical applications
of  PACT \cite{gamelin2009real,li2010real,buehler2010video,xiang20134,lou2016impact}, 
where the sought-after initial pressure distribution can depend on time denoted as $p_0(\mathbf r,t)$.
Spatiotemporal, or dynamic,  image reconstruction methods for PACT
 seek to reconstruct a sequence of images that correspond to a
 collection of temporal samples.
 In a dynamic PACT study, the measurement data collected
at a particular time point is referred to as a data frame.
 It is assumed that $p_0(\mathbf r)$ remains static during the
 acquisition of each data frame. This condition can approximately be
satisfied if the temporal resolution of the imaging system is sufficiently high,
such as when fixed arrays of transducers are employed.

A simple approach to dynamic reconstruction is to simply
employ a (static) conventional PACT reconstruction method to reconstruct
estimates of $p_0(\mathbf r,t)$ corresponding to times $t$ at which the data
frames were acquired.
  Such approaches are referred to as frame-by-frame reconstruction (FBFIR)
 methods~\cite{li2010real,buehler2010video,xiang20134,chatni2012tumor}.
 A limitation of  FBFIR methods is that they 
do not exploit statistical correlations between data frames and
 are therefore computationally and
 statistically suboptimal~\cite{wernick1999fast,tsao2003k,lingala2011accelerated}.
Below, a few representative examples of  spatiotemporal image reconstruction strategies for
 dynamic PACT that leverage correlations between data frames are reviewed.

 Unlike the FBFIR methods, spatiotemporal image reconstruction (STIR) methods for dynamic PACT jointly reconstruct the sequence of object frame estimates by using all of the data frames.
One such method that has received attention is the low rank matrix estimation (LRME)-STIR method~\cite{wang2014fast,tremoulheac2014dynamic}. The LRME-STIR method exploits the fact that, for many dynamic PACT applications, the image sequence can be approximately described as a low-rank matrix whose rank is typically much smaller than the number of temporal samples. Hence, the LRME-STIR method consists of applying a data denoising step followed by
 image reconstruction conducted in the domain of the singular system
 of the low rank data matrix.  The performance of the LRME-STIR method was compared with that of conventional FBFIR method with both computer-simulated and 2D experimental data. The results of the study demonstrated that the LRME-STIR method was not only computationally more efficient but also yielded more accurate dynamic PACT images than conventional FBFIR methods~\cite{wang2014fast}. 

 Recently, to improve upon low-rank approximation-based methods,
 a generalized spatio-temporal modeling framework has been
proposed that can encode information about a wide range of dynamics~\cite{lucka2018enhancing}. In this approach, the image dynamics are described by an explicit partial
differential equation (PDE) model chosen \emph{a priori} such that it reflects the underlying dynamics.  Subsequently, the image sequence and the
 corresponding motion parameters of the PDE are jointly estimated by minimizing a  variational energy cost function. In particular, the motion model employed was based on the popular optical flow equation. The approach was applied successfully to a simulated data as well as to a challenging 3D scenario with experimental data. The results of the study showed that both the simulated and the experimental data-based reconstructions showed a significant improvement in image quality over FBFIR reconstructions~\cite{lucka2018enhancing}. Furthermore, the reconstructed motion fields, or the parameters of the PDE provided additional information regarding the
 dynamics.   For the case where the temporal dynamics are relatively simple,
 low-dimensional parametric models can also be effectively be used to constrain the
dynamic  image reconstruction problem~\cite{chung2017motion}.

In general, there remains an important need for the development of reconstruction
methods for dynamic PACT that are computationally efficient and can compensate for
important physical factors.  For example, there have been no reported methods for dynamic 
PACT that can compensate for the effects of unknown heterogeneities in the acoustic
properties of an object.

\subsection{PACT in the presence of elastic media}


Conventional PACT reconstruction methods assume that the to-be-imaged
object can be described as a fluid acoustic medium.
When only soft biological tissues are present, this assumption is well-accepted.
However, calcified tissues or bone are more accurately described as an elastic, or solid,
medium.  In elastic media, the propagation of
  both longitudinal pressure waves as well as transverse shear waves are supported,
each having a distinct speed of propagation~\cite{acousticskull}.
As such, to preserve image quality, PACT reconstruction methods  for use in
applications in which elastic solids are present need to be predicated upon the elastic wave
equation as described below. 

Let the \mbox{photoacoustically-induced} stress tensor at location $\mathbf{r} \in \mathbb{R}^3$ and time $t \geq 0$ be defined as
\begin{align} 
\boldsymbol{\sigma}(\mathbf{r},t)\equiv
\begin{bmatrix}
&\boldsymbol{\sigma}^{11}(\mathbf{r},t) &\boldsymbol{\sigma}^{12}(\mathbf{r},t) &\boldsymbol{\sigma}^{13}(\mathbf{r},t)\hspace{10pt}
\\ 
&\boldsymbol{\sigma}^{21}(\mathbf{r},t) &\boldsymbol{\sigma}^{22}(\mathbf{r},t) &\boldsymbol{\sigma}^{23}(\mathbf{r},t)\hspace{10pt}
\\
&\boldsymbol{\sigma}^{31}(\mathbf{r},t) &\boldsymbol{\sigma}^{32}(\mathbf{r},t) &\boldsymbol{\sigma}^{33}(\mathbf{r},t)\hspace{10pt} 
\end{bmatrix},
\end{align} 
where $\boldsymbol{\sigma}^{ij}(\mathbf{r},t)$ represents the stress in the $i^{\text{th}}$ direction acting on a plane perpendicular to the $j^{\text{th}}$ direction. Additionally, let $p_0(\mathbf{r})$ denote the photoacoustically-induced initial pressure distribution  within the object, and  $\dot{\mathbf{u}}(\mathbf{r}, t)$ $\equiv$ ($\dot{u}^1(\mathbf{r},t), \dot{u}^2(\mathbf{r},t), \dot{u}^3(\mathbf{r},t)$) represent the vector-valued acoustic particle velocity. Let $\rho(\mathbf{r})$ denote medium's density distribution and $\lambda(\mathbf{r})$, $\mu(\mathbf{r})$  represent the Lam\'e parameters that describe the full elastic tensor of the linear isotropic media.

The pressure  and shear wave propagation speeds are given by, 
\begin{subequations} 
	\begin{align}
	c_l(\mathbf{r}) = \sqrt{\frac{\lambda(\mathbf{r}) + 2 \mu(\mathbf{r})}{\rho(\mathbf{r})}} \text{ and }
	c_s(\mathbf{r}) = \sqrt{\frac{\mu(\mathbf{r})}{\rho\left(\mathbf{r}\right)	}},
	\end{align}
\end{subequations}
respectively.
In a 3D heterogeneous linear isotropic elastic medium with an acoustic absorption coefficient $\alpha(\mathbf{r})$, the propagation of $\dot{\mathbf{u}}(\mathbf{r},t)$ and $\boldsymbol{\sigma}(\mathbf{r},t)$ can be modeled by the following two coupled equations~\cite{boore1972finite,alterman1968propagation,madariaga1998modeling,virieux1986p}:
\begin{subequations}
	\begin{equation}
	\partial_t \dot{\mathbf{u}}\left(\mathbf{r},t\right) + \alpha\left(\mathbf{r}\right) \dot{\mathbf{u}}\left(\mathbf{r},t\right)= \frac{1}{\rho\left(\mathbf{r}\right)}\Big(\nabla \cdot \boldsymbol{\sigma}\left(\mathbf{r},t\right)\Big)
	\label{eq:subeq1}
	\end{equation}
	and
	\begin{equation}
	\partial_t \boldsymbol{\sigma}\left(\mathbf{r},t\right) = \lambda(\mathbf{r}) \textbf{tr}(\nabla\dot{\mathbf{u}}\left(\mathbf{r},t\right)) \mathbf{I} +  \mu(\mathbf{r})(\nabla \dot{\mathbf{u}} \left(\mathbf{r},t\right) + \nabla \dot{\mathbf{u}}\left(\mathbf{r},t\right) ^ {T}),
	\label{eq:subeq2}
	\end{equation}
	\text{subject to the initial conditions}
	\begin{align}\label{eq:Ini}
	\boldsymbol{\sigma}_0(\mathbf{r}) \equiv \boldsymbol{\sigma}(\mathbf{r},t)|_{t = 0} = -\frac{1}{3} p_0(\mathbf{r})\mathbf{I}, \ \ \ \ \ \ \  \dot{\mathbf{u}}\left(\mathbf{r},t\right)|_{t=0} = 0.
	\end{align}
	\label{eq:Elastic}
\end{subequations} 
Here, $\textbf{tr}$ $(\cdot)$ is the operator that calculates the trace of a matrix and $\mathbf{I}\in \mathbb{R}^{3 \times 3}$ is the identity matrix.
In Eqn.~\eref{eq:Ini}, it has been assumed that the initial
pressure distribution  $p_0(\mathbf{r})$ is compactly supported in a fluid medium where the shear modulus $\mu(\mathbf{r}) = 0 $.  In transcranial PACT \cite{huang2012aberration}, for example, this corresponds to the situation
where the initial photoacoustic wavefield is produced within
the soft tissue enclosed by the skull. Equation (\ref{eq:Elastic}) represents
a generalization (but with a diffusive instead of power law
 attenuation model) of the canonical PACT forward model for fluid media
 in Eqn.\ (\ref{eq:2a_loss}).

For the case of layered elastic media, an analytic image reconstruction formula
has been established \cite{schoonover2011compensation}. More generally, a 
methodology for establishing a D-D forward model based on Eqn.\ (\ref{eq:Elastic})
and the associated matched adjoint operator has been established  and investigated
within the context of transcranial PACT \cite{mitsuhashi2017forward}.
The discretization of the elastic wave equation was based on the finite difference in
time domain and space, which facilitated large scale parallelization of
the forward and adjoint operators using multiple GPUs.
By use of the D-D operators, a variety of optimization-based image reconstruction
methods can be employed \cite{poudel2017iterative}.
An alternative formulation of the D-D forward and adjoint operators
that was established by use of  the k-space pseudo-spectral method has also
been reported \cite{javaherian2018continuous}.

There remains an important need for the development of reconstruction
methods for transcranial PACT that are computationally efficient and can compensate for
important physical factors.  For example, there have been no reported methods for 
transcranial PACT that can compensate for the effects of unknown acoustic heterogeneities
within the skull of a subject.

\section{Summary}

\label{sec:section8}

In this topical review,  a survey of the acoustic
inverse problem in PACT has been presented.  Specifically,    
 the PACT image reconstruction problem of estimating the photoacoustically-induced
initial pressure distribution has been reviewed
 within the context of modern optimization-based image reconstruction methodologies.
 Imaging models that relate the measured photoacoustic wavefields to the sought-after
 initial pressure distribution were described in their continuous and discrete forms.
Subsequently, a  description of how important physical factors relevant to PACT can be incorporated
into the imaging models was also provided.

Details regarding the implementation of optimization-based PACT reconstruction methods that are not widely
available in other review papers were presented.
  For example, descriptions of  the implementations of different discrete adjoint operators
employed in PACT were provided. Such adjoint operators are a key component
of gradient-based algorithms that are employed to solve optimization-based reconstruction problems.
  Additionally, non-conventional formulations
 of the PACT image reconstruction problem were also reviewed, in which the
acoustic parameters of the medium are concurrently estimated along with the
initial pressure distribution.

\ack     
The authors would like to thank Dr. Thomas P. Matthews, Dr. Kun Wang and Dr. Chao Huang for their insightful discussions regarding 3D PACT. The authors would also like to thank Dr. Umberto Villa and the anonymous reviewers for their careful reading of the manuscript. This work was supported in part by awards NSF 1614305 and NIH NS102213. 
\section*{References}
\bibliographystyle{jphysicsB}
\bibliography{reflect_JP,reflect_YL,reflect2,reflect_Jin}   

\clearpage
\newpage

\end{document}